\documentclass[structabstract]{aa}

\usepackage{amsmath}
\usepackage{txfonts}
\usepackage{booktabs}
\usepackage{array}
\usepackage{graphicx}
\usepackage{natbib,twoopt}
\usepackage{upgreek}
\usepackage{color}
\usepackage{xcolor}
\usepackage{wasysym}
\usepackage[normalem]{ulem}
\usepackage[colorlinks,linkcolor=blue,anchorcolor=blue,citecolor=blue,breaklinks=true]{hyperref}
\usepackage{tablefootnote}
\usepackage{subcaption}
\bibpunct{(}{)}{;}{a}{}{,} %% natbib format like A&A and ApJ
\begin{document}

\title{A spatially resolved radio spectral study of the galaxy \object{M\,51} }
\author{L. Gajovi\'c\inst{1} \and 
B. Adebahr\inst{2,3} \and 
A. Basu\inst{4,5} \and
V. Heesen\inst{1} \and
M. Br\"uggen\inst{1} \and
F. de Gasperin\inst{6} \and
M. A. Lara-Lopez\inst{7} \and
J. B. R. Oonk\inst{8,3} \and
H. W. Edler\inst{1} \and
D. J. Bomans\inst{2} \and
R. Paladino\inst{6} \and
L. E. Gardu\~no\inst{9} \and
O. L\'opez-Cruz\inst{9} \and
M. Stein\inst{2} \and
J. Fritz\inst{10} \and
J. Piotrowska\inst{11} \and
A. Sinha\inst{4}}
\institute{Hamburger Sternwarte, University of Hamburg, Gojenbergsweg 112, 21029 Hamburg, Germany
\and Ruhr University Bochum, Faculty of Physics and Astronomy, Astronomical Institute (AIRUB), Universit\"atsstrasse 150, 44801 Bochum, Germany 
\and ASTRON, PO Box 2, 7990 AA Dwingeloo, The Netherlands
\and Th\"uringer Landessternwarte, Sternwarte 5, 07778 Tautenburg, Germany
\and Max-Planck-Institut f\"ur Radioastronomie, Auf dem H\"ugel 69, 53121 Bonn, Germany
\and INAF - Istituto di Radioastronomia, via P. Gobetti 101, 40129 Bologna, Italy
\and Departamento de F\'{i}sica de la Tierra y Astrof\'{i}sica, Instituto de F\'{i}sica de Part\'{i}culas y del Cosmos, IPARCOS.
Universidad Complutense   de Madrid (UCM), E-28040, Madrid, Spain
\and Leiden Observatory, Leiden University, PO Box 9513, NL-2300 RA Leiden, The Netherlands
\and Instituto Nacional de Astrof{\'\i}sica, \'Optica y Electr\'onica (INAOE), Coor. de Astrof{\'\i}sica, Luis Enrique Erro No.1,  Tonantzintla, Puebla, M\'exico C.P. 72840.
\and  Instituto de Radioastronom\'ia y Astrof\'isica, UNAM, Campus Morelia, A.P. 3-72, C.P. 58089, Mexico
\and Astronomical Observatory of the Jagiellonian University, ul.
Orla 171, 30-244 Krak\'ow, Poland
}

\date{Received - / Accepted -}

\abstract
{Radio continuum emission from galaxies at gigahertz frequencies can be used as an extinction-free tracer of star formation. However, at frequencies of a few hundred megahertz, there is evidence for low-frequency spectral flattening.}
{We wish to better understand the origin of this low-frequency flattening and, to this end, perform a spatially resolved study of the nearby spiral galaxy M\,51. We explore the different effects that can cause flattening of the spectrum towards lower frequencies, such as free--free absorption and cosmic-ray ionisation losses.}
{We used radio continuum intensity maps between 54 and 8350\,MHz at eight different frequencies, with observations at 240\,MHz from the Giant Metrewave Radio Telescope presented for the first time. We corrected for contribution from thermal free--free emission using an H\,$\upalpha$ map that has been extinction-corrected 
with 24\,$\upmu$m data. We fitted free--free absorption models to the radio spectra to determine the emission measure (EM) as well as polynomial functions to measure the non-thermal spectral curvature. We also obtained a new extinction-corrected H\,$\upalpha$ intensity map from the Metal-THINGS survey using integral field unit spectroscopy.}
{The non-thermal low-frequency radio continuum spectrum between 54 and 144\,MHz is very flat and even partially inverted, particularly in the spiral arms; contrary, the spectrum at higher frequencies shows the typical non-thermal radio continuum spectrum. However, we do not find any correlation between the EMs calculated from radio and from H\,$\upalpha$ observations; instead, the non-thermal spectral curvature weakly correlates with the H\,{\sc i} gas mass surface density. This suggests that cosmic-ray ionisation losses play an important role in the low-frequency spectral flattening.}
{The observed spectral flattening towards low frequencies in M\,51 is caused by a combination of ionisation losses and free--free absorption.  The reasons for this flattening need to be understood in order to use sub-GHz frequencies as a star-formation tracer.}


\maketitle

\section{Introduction}

At radio frequencies of a few gigahertz and below, the majority of continuum emission from galaxies is non-thermal synchrotron radiation produced by cosmic-ray electrons \citep{1997A&A...322...19N, 2012MNRAS.419.1136B, 2017ApJ...836..185T}. These cosmic rays are accelerated in the shock-front of supernovae explosions, and therefore they originate from regions with higher star formation \citep{2014BASI...42...47G}. The integrated spectrum of nearby galaxies at gigahertz frequencies (approximately $0.5$--$10$\,GHz) mostly follows a power law where the flux density scales with frequency as $S_\nu \propto \nu^\alpha$; the radio spectral index $\alpha$ has values between $-1.2$ and $-0.5$ \citep{1982A&A...116..164G,Li2016,2017ApJ...836..185T}. If frequencies below approximately 300\,MHz are included though, the observed spectra are curved with the spectral curvature of $\beta=-0.2$, which denotes the change in spectral index per logarithmic frequency decade assuming constant curvature \citep{2015AJ....149...32M}. 

While the value for the radio spectral index generally agrees with models of the overall injection spectrum of cosmic rays from supernovae \citep[$\upalpha\approx-0.5$;][]{1987PhR...154....1B}, including their energy losses mainly due to synchrotron and inverse Compton radiation while propagating away from their origin \citep{1982A&A...116..164G,2013pss5.book..641B,2017ARA&A..55..111H}, the origin of the curvature is still a matter of debate \citep{2015MNRAS.449.3879B,2018A&A...615A..98M}. \citet{1990A&A...239..424P} and \citet{1991A&A...250..302P,1991A&A...252..493P} provide an explanation for curved spectra via a combination of ionisation and electronic excitation \citep{1975ApJ...196..689G}, inverse Compton radiation, synchrotron radiation, and relativistic bremsstrahlung, which would cause a smooth change in the spectral index of $\Delta\alpha=0.5$ from higher to lower radio frequencies. On the other hand, \citet{1990ApJ...352...30I} observed a dependence of the curvature with the inclination of the galaxies and concluded that a fragmented cool ionised medium with temperatures of $500\textrm{--}1000$\,K is responsible for the curvature by thermally absorbing a part of the emitted synchrotron emission at lower frequencies. However, \citet{1991A&A...251..442H} used the same data as \citet{1990ApJ...352...30I} and could not find any correlation between the curvature in the sampled galaxies and their inclinations. A more recent study by \citet{2018A&A...619A..36C} using additional data from the LOFAR Multifrequency Snapshot Sky Survey (MSSS) at 150\,MHz \citep{2015A&A...582A.123H} confirmed the results by \citet{1991A&A...251..442H}.

Furthermore, integrated spectra of star-forming galaxies encompass a mixture of the complex interplay between thermal and non-thermal components and different energy loss and propagation effects in the most likely non-isotropic and inhomogeneous medium \citep{2015MNRAS.449.3879B,2000A&A...354..423L}. Studies of large samples of these objects are therefore limited by obtaining high-quality, high-resolution observations at low radio frequencies ($\lesssim$$300$\,MHz). Advancements in data calibration and imaging techniques at these low-frequencies, especially for radio interferometers such as the LOw Frequency ARray \citep[LOFAR;][]{2014A&A...566A.127T,2018A&A...611A..87T} and the Giant Metrewave Telescope  \citep[GMRT;][]{2009A&A...501.1185I,2014ASInC..13..469I,2017A&A...598A..78I}, now allow the production of high dynamic range images with resolutions matching those of observations performed at gigahertz frequencies.

Spatially resolved studies are currently limited to nearby objects where individual star-forming regions can be analysed, and the radio emission of spiral arm, inter-arm and surrounding regions can be separated. \citet{2006JPhCS..54..156R} showed that thermal absorption plays a key role in shaping the spectra of the centre of the Milky Way below 500\,MHz while its integrated-spectrum turns over at about 3\,MHz \citep{1973ApJ...180..359B}. Turnovers have also been detected for individual sources in the starburst galaxy M\,82 \citep{2017A&A...608A..29A} and the integrated flux densities for the core regions of M\,82 \citep{2013A&A...555A..23A} and Arp\,220 \citep{2016A&A...593A..86V} at around 1000\,MHz. For these two objects a flattening was also observed for the integrated (global) radio spectra \citep{1988A&A...190...41K,1992ARA&A..30..575C,2000ApJ...537..613A}, just like in the starburst galaxy NGC\,253 \citep{2015AJ....149...32M}. Several of the compact H\,{\sc ii}-regions in the dwarf galaxy IC\,10 show evidence of thermal free--free absorption in the radio spectra of the 320\,MHz observations \citep{2017MNRAS.471..337B}. Recently, evidence of low frequency absorption was also found in region of the edge-on galaxy NGC\,4631 with a very flat radio spectrum \citep{2023A&A...670A.158S}. Local spectral turnovers were also observed in jellyfish galaxies, possibly due to gas compression and subsequent ionization losses \citep{2022ApJ...934..170L, 2022ApJ...924...64I,2023arXiv231020417R}.

\begin{table}
	\caption{Basic properties of M51.}
	\label{table_overview}
	\centering
	\begin{tabular}{@{} lc @{}}
		\toprule
		\toprule
		Name & M\,51 \\
		Alternative names & NGC\,5194/5155 \\
		RA$_{2000}$ & 13 \fh29 \fm56 \fs 2 \\
		DEC$_{2000}$ & +47\degr13\arcmin50\arcsec \\
		Type & Galaxy Pair (Sa + Sc)$^\text{a}$ \\
		Apparent size (D$_{25}$) & $9\arcmin$ \\
		Distance & $8.58\pm0.10$\,\text{Mpc}$^\text{b}$ \\
		Position angle & 12\fdg 0$^\text{c}$ \\
		Inclination angle & $-$20\fdg 3$^\text{c}$ \\
		\bottomrule
	\end{tabular}
	\begin{list}{}{}
		\item[References:]$^\text{a}$\citet{1985BICDS..29...87K}, $^\text{b}$\citet{2016ApJ...826...21M}, $^\text{c}$\citet{2013ApJ...762L..27H}
	\end{list}
\end{table}

We choose M\,51 as the target primarily as it benefits from multi-frequency radio continuum data. Also, the face-on orientation and the proximity are important because they make it easier to resolve and separate the spiral arms. Other properties of M\,51 are listed in Table \ref{table_overview}. The entire galaxy is now mapped at 54 and 144\,MHz with LOFAR \citep{2021A&A...648A.104D,2022A&A...659A...1S}. Furthermore, there are data at 1370 and 1699\,MHz taken with the Westerbork Synthesis Radio Telescope \citep[WSRT;][]{2007A&A...461..455B}. At even higher frequencies, \citet{2011MNRAS.412.2396F} presented maps taken with the Very Large Array (VLA) at 4850 and 8350\,MHz. VLA maps were combined with maps observed with the 100-m Effelsberg telescope in order to observe all angular scales. These data were analysed by \citet{2023A&A...672A..21H} who studied the transport of cosmic-ray electrons at kpc-scales and found it to be energy-independent diffusion for electrons with energies below 10\,GeV.

In this paper we extend the data with a new 240\,MHz map from the Giant Metre Radio Telescope. We used eight data-sets observed at five different radio facilities (LOFAR, GMRT, WSRT, VLA, Effelsberg) over the frequency range 54--8350\,MHz to perform a spatially resolved radio spectral study of the nearby spiral galaxy M\,51. For the first time we were able to separate the actively star-forming and non star-forming regions of a nearby galaxy down to frequencies as low as 54\,MHz. This allows us to investigate the relation between star-formation and the flattening of the radio spectrum, as well as to perform an examination of the origin of the flattening.

This article is structured as follows. In Sect.\,\ref{section_data} we present our dataset spanning nine frequency windows. Section \ref{section_analysis} describes how we conduct the main data analysis which includes subtraction of thermal emission (Sect.\,\ref{section_thermal_separation}), spectral index maps (Sect.\,\ref{section_SI_SC}), separation of galaxy regions (Sect.\,\ref{section_spiral_inter-arm}), and comparison of low- and high-frequency spectral indices (Sect.\,\ref{section_low_high}). Spectral index flattening and low-frequency turnovers are investigated in Sect.\,\ref{section_turnover}. We summarize and conclude in Sect.\,\ref{section_discussion}.

\section{Data handling}
\label{section_data}

\subsection{Giant Metrewave Radio Telescope observations}

\begin{figure}
\centering
	\resizebox{\hsize}{!}{\includegraphics{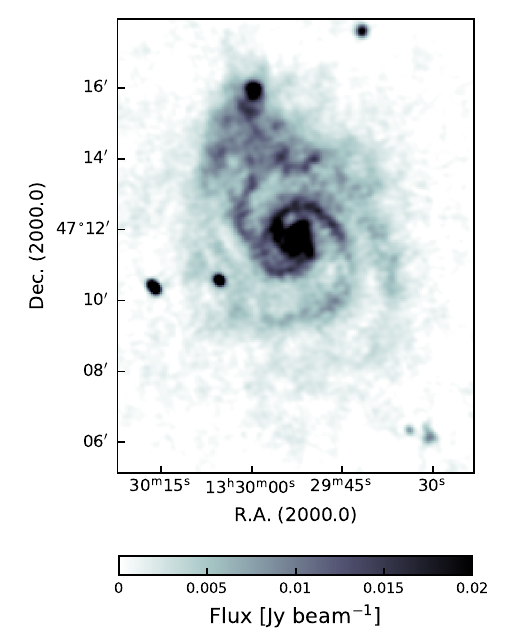}}
	\caption{Radio continuum map of M\,51 at 240\,MHz observed with the GMRT. The map has an angular resolution of $14\farcs 0\times 12\farcs 2$ and an rms noise of $500\,\upmu$Jy\,beam$^{-1}$.}
	\label{image_240MHz}
\end{figure}

We have analyzed archival, so far not published, data of M\,51 at 240\,MHz, observed with the Giant Metrewave Radio Telescope \citep[GMRT;][]{1991CSci...60...95S} using the old hardware correlator (project code: 10AFA01). The map is presented here for the first time, see Fig.~\ref{image_240MHz}. The data were recorded by splitting the 16\,MHz bandwidth into two bands of 8\,MHz each covered using 64 125-kHz wide channels, the upper and lower sidebands, USB and LSB, respectively. We have analyzed only the USB data centered at 241.25\,MHz because the LSB data were unusable due to radio frequency interference (RFI) and correlator-related errors. The total duration of these observations is 11\,h, wherein 4\,min scans on a nearby, bright, point source, 3C\,286, were interleaved every 25\,min. 3C\,286 was used to calibrate the flux scale, bandpass and phase. Additional 15\,min scans on 3C\,286 were performed at the begining and end of the observations run. In these observations, a total of 8.3\,h was spent on M\,51. The data were analyzed in \texttt{AIPS}\footnote{\texttt{AIPS}, the Astronomical Image Processing System, is free software available from the National Radio Astronomy Observatory.} using the standard analysis protocol at low frequencies. Data were manually inspected, and data affected by RFI and bad baselines were flagged using the \texttt{TVFLG, WIPER and SPFLG} tasks. The task \texttt{FLGIT} was used to automatically remove RFI-affected frequency channels that have residual spectral flux above $4\sigma$ level. Overall, in these observations, less than $\sim35\%$ of the data were found to be corrupted in the USB, including three non-working antennas and corrupted baselines. We used the \citet{1977A&A....61...99B} absolute flux density scale to determine the flux density of $29.5 \pm 1.6$\,Jy at 241.25\,MHz for 3C\,286.\footnote{Note that, the flux density of 3C\,286 obtained with the \citet{1977A&A....61...99B} is consistent with 28.15\,Jy from the \citet{2017ApJS..230....7P} absolute scale within the error.} RFI-removal and determining gain solutions were iteratively performed, and when the closure-phase errors determined using 3C\,286 were below 1\%, solutions were applied to the target M\,51.

The task \texttt{IMAGR} was used to deconvolve the calibrated data. In order to minimize the effect of wide-field imaging with non-coplanar baselines, we used polyhedron imaging \citep{1992A&A...261..353C} by subdividing the field of view ($\approx\!3.7^\circ$) into $9\times9 = 81$ smaller facets. After flagging the first and the last two channels, we vector-averaged four adjacent frequency channels of 125\,kHz each into 15 500-kHz-channels to ensure that the bandwidth smearing near 240\,MHz is less than the size of the synthesized beam. Thus, the final image was made using 7.5\,MHz bandwidth. Five rounds of phase-only self-calibration were performed by choosing point sources above $8\sigma$ using the task \texttt{CCSEL}. The solution interval was progressively reduced during each round, starting from 3\,min at the start to 0.5\,min in the last round. Additional flagging were also done after each round until in the fifth round, the closure-phase error was below 0.5\%. To prepare the final image, we employed the \texttt{SDI clean} algorithm \citep{1984A&A...137..159S} to deconvolve the extended emission in M\,51. After ensuring that point-like emission were cleaned using \texttt{BGC clean} algorithm, \texttt{SDI clean} was used, and the clean-box masks were manually changed in order to deconvolve different scales.
The final image at 240\,MHz used in this work was produced at an angular resolution of $14\farcs 0 \times 12\farcs 2$ with an rms noise of $\approx$$500\,\upmu$Jy\,beam$^{-1}$. While not a very sensitive map, it suffices for our analysis. The galaxy-integrated flux density of M\,51 at 240\,MHz within the $3\sigma$ contour is found to be $6.07 \pm 0.63$\,Jy. As can be seen in Fig.~\ref{plot_integrated}, the total flux density of M\,51 agrees well with those in the literature when interpolated from higher and lower frequency observations.

The uncertainties in the estimated flux density depend on the absolute flux scale error and on the errors associated with uncalibrated system temperature  ($T_{\rm sys}$) variations \citep{basu12}. The absolute flux scale error for 3C\,286 is found to be 5.4\% \citep{1977A&A....61...99B}, and the $T_{\rm sys}$ variation for the old GMRT system, as used for the observations in this paper, was estimated to be $\sim5\%$ \citep{2004MNRAS.349L..25R}. So, the overall systematic error, added in quadrature, is expected to be $\sim7.5\%$. We use a slightly more conservative value of 10\% for the systematic flux error of the GMRT data. The statistical measurement error due to the rms noise in the map is added on top of that, but that contribution is relatively small.

\subsection{Archival radio continuum observations}

In total, radio continuum image data of M\,51 were collected in nine different frequency bands between 54\,MHz and 8350\,MHz (see Table \ref{table_data}). All retrieved images were inspected for artefacts and their integrated fluxes cross-checked with literature values. Preferentially single-dish or compact interferometric measurements were used to ensure that our images are not affected by the missing spacing problem. 

\begin{table*}
\begin{minipage}{\textwidth}
	\caption{Observation parameters for the nine radio maps used in the analysis.}
	\label{table_data}
	\centering
	\newcolumntype{0}{>{\centering\arraybackslash} m{2.5cm} }
	\newcolumntype{1}{>{\centering\arraybackslash} m{1.8cm} }
	\newcolumntype{2}{>{\centering\arraybackslash} m{2.0cm} }
	\newcolumntype{3}{>{\centering\arraybackslash} m{1.8cm} }
	\newcolumntype{4}{>{\centering\arraybackslash} m{1.8cm} }
	\newcolumntype{5}{>{\centering\arraybackslash} m{1.5cm} }
	\newcolumntype{6}{>{\centering\arraybackslash} m{1.5cm} }
	\renewcommand{\arraystretch}{1.5}
	\begin{tabular}{@{} 0 1 2 3 4 5 6 @{}}
		\toprule
		\toprule
		Telescope (Survey) & Central frequency [MHz] & Beam size & $\sigma$ rms noise [$\upmu$Jy] & convolved $\sigma_c$ rms noise [$\upmu$Jy] & Flux uncertainty $\varepsilon_\nu$ [\%] & Reference \\
		\midrule
        LOFAR (LoLSS) & 54 & $15\farcs 0\times 15\farcs 0$ & 1500 & 1500 & 10 & 1 \\
        LOFAR (LoTSS) & 144 & $6\farcs 0\times 6\farcs 0$ & 45 & 100 & 10 & 2 \\
		GMRT & 240 & $14\farcs 0\times 12\farcs 2$ & 500 & 850 & 10 & 3 \\
		GMRT & 333 & $15\farcs 0\times 15\farcs 0$ & 200 & 350 & 10 & 4 \\
		GMRT \footnote{Map not used in the analysis due to possible deconvolution errors.} & 619 & $5\farcs 7\times 4\farcs 6$ & 27 & 125 & 10 & 5 \\
		WSRT (SINGS) & 1370 & $17\farcs 5\times 12\farcs 5$ & 22 & 25 & 5 & 6 \\
		WSRT (SINGS) & 1699 & $13\farcs 5\times 10\farcs 0$ & 27 & 30 & 5 & 6 \\
		VLA + Effelsberg & 4850 & $15\farcs 0\times 15\farcs 0$ & 30 & 30 & 5 & 7 \\
		VLA + Effelsberg & 8350 & $15\farcs 0\times 15\farcs 0$ & 20 & 20 & 5 & 7 \\
		\bottomrule
	\end{tabular}
	\begin{list}{}{}
		\item[References:] 1) \citet{2021A&A...648A.104D}, 2) \citet{2022A&A...659A...1S}, 3) \text{This work}, 4) \citet{2016A&A...592A.123M},
        5) \citet{2013arXiv1309.4646F},
        6) \citet{2007A&A...461..455B}, 7) \citet{2011MNRAS.412.2396F}
	\end{list}
\end{minipage}
\end{table*}

The lowest frequency image at 54\,MHz is from the LOFAR LBA Sky Survey \citep[LoLSS;][]{2021A&A...648A.104D} using the low-band antennae. Flux calibrators were used to calibrate direction-independent effects as well as the bandpass response of the instrument. Additionally, the flux density of sources in the survey was compared to the Rees survey \citep[8C;][]{1995MNRAS.274..447H} at 38\,MHz, the Very Large Array (VLA) Low-Frequency Sky Survey redux \citep[VLSSr;][]{2012RaSc...47.0K04L} at 74\,MHz, the LOFAR Two-metre Sky Survey data release 2 \citep[LoTSS-DR2;][]{2022A&A...659A...1S} at 144\,MHz, and the NRAO VLA Sky Survey \citep[NVSS;][]{1998AJ....115.1693C} at 1400\,MHz. A conclusive estimate of the flux density accuracy could not be derived, but it is suggested that assuming a conservative 10\,\% error on the LoLSS flux density scale is beneficial \citep{2021A&A...648A.104D}. \citet{2022A&A...664A..83H} investigated 45 nearby galaxies in LoTSS-DR2 including M\,51, we use their re-processed map of M51. They confirmed that the integrated flux densities of the $6\arcsec$ and $20\arcsec$ maps are identical for the same integration area. This ensures that the high-resolution map, which we are using, is sufficiently deconvolved. The flux densities in the work by \citet{2022A&A...664A..83H} were matched with the LoTSS-DR2 scale which is has an error below 10\% \citep{2022A&A...659A...1S}. Therefore, in this work we assume a flux uncertainty of 10\% for the LoTSS map.

The GMRT map at 333\,MHz was already presented by \citet{2016A&A...592A.123M}. They have used it for an analysis of the large-scale diffuse emission of M\,51 and found no issues in relation to missing emission for the inner part of the disc. In contrast, the 619\,MHz GMRT image of \citet{2013arXiv1309.4646F} showed a `negative bowl' around the diffuse emission of M\,51 indicating deconvolution errors arising from missing emission. We have, therefore, refrained from using the diffuse emission for further quantitative analyses. However, we believe that the bright emissions on small-scales, especially in the inner parts of the galaxy to be less affected from missing flux issues.  The WSRT images at 1370 and 1699\,MHz from \citet{2007A&A...461..455B} were cross-checked with the VLA image at 1400\,MHz published in \citet{2011MNRAS.412.2396F} and with literature flux density values. No differences in flux densities or morphology were found. The WSRT-SINGS observations were bracketed by observations of the total intensity calibration sources yielding an absolute flux density calibration accuracy better than 5\% \cite{2007A&A...461..455B}. We therefore assume 5\% flux uncertainty for the WSRT maps used in this work. Finally, the VLA images at 4850 and 8350\,MHz have already been combined with single-dish Effelsberg telescope data as described in \citet{2011MNRAS.412.2396F}, so that we do not expect any missing flux. Flux measurement errors are usually assumed to be 5\% for VLA data \citet{2013MNRAS.430.2137K}.

We convolved all images to the largest common beam of $17\farcs 5\times 15\farcs 0$ (see Table~\ref{table_data}) using the routine \texttt{CONVOL} of the Multichannel Image Reconstruction, Image Analysis and Display  \citep[\texttt{MIRIAD},][]{1995ASPC...77..433S} software package. All images were then aligned to a common world coordinate system,  re-gridded to a pixel size of $3\arcsec$, and transformed to the  reference pixel and image size. This was achieved using the \texttt{REGRID} routine of \texttt{MIRIAD}. The rms noise, $\sigma$, of each individual image was determined by calculating the standard deviation over an emission-free area. For further analysis, only pixels above $4.5\,\sigma$ were considered. 

We assumed an absolute flux scale uncertainty of 10\,\% for the low frequency data (for LOFAR and GMRT) and 5\,\% for all other data (see Table \ref{table_data}). These estimates incorporate all errors and are conservative. There are several contributions, first there is the accuracy of the absolute flux density of the calibrator models. This is 3--5\,\% \citep{Perley_17a}. However, for spectral index studies like our work the relative scale is more relevant, the uncertainty of which is around 1\,\%. Deconvolution errors are around 1\,\% but increase at low signal-to-noise ratios \citep{Offringa_14a}. The background noise can be neglected except for local measurements, where we add the rms noise in quadrature. Other errors such as the uncertainty of primary beam correction can be also neglected as the primary beam size is much larger than the size of source. For the combined VLA and Effelsberg maps the uncertainty is limited by the feathering procedure which allows one to combine the interferometric and single-dish images \citep{Cotton2017}. Their uncertainty may be higher in areas of low signal-to-noise ratios, but we restrict most of our analysis to areas where this should make no strong difference.

\subsection{Integral field unit spectroscopy}
\label{section_spectroscopy}

\begin{figure}
\centering
	\resizebox{\hsize}{!}{\includegraphics{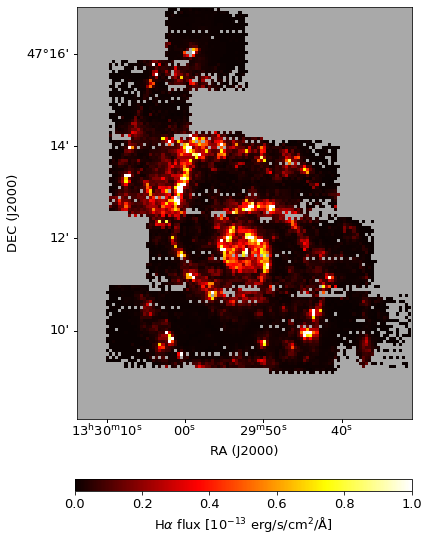}}
	\caption{ H\,$\upalpha$ flux density map of M51 from the Metal-THINGS survey corrected for extinction using the Balmer decrement. The gray pixels were not observed. }
	\label{image_IFU}
\end{figure}

We complement our radio continuum data with new optical integral field unit spectroscopy data from the
Metal-THINGS survey \citep{2021ApJ...906...42L}. 
Metal-THINGS is a survey of nearby galaxies observing with the George Mitchell and Cynthia Spectrograph (GCMS; formerly known as VIRUS-P), mounted to the 2.7-m Harlan J.\ Smith telescope located at the McDonald Observatory in Texas. The integral field unit (IFU) has a field of view of $100\arcsec \times 102\arcsec$. Due to the large angular size of M\,51, a total of 12 pointings were needed to cover the entire galaxy. The spectroscopic data were processed with the spectral synthesis code \texttt{STARLIGHT} \citep{Cid2005} as described in \citet{2021ApJ...906...42L, 2023A&A...669A..25L}. We used an H\,$\upalpha$ flux map corrected for extinction, where the correction was performed using the Balmer decrement. The map is shown in figure \ref{image_IFU}.

\section{Data analysis}
\label{section_analysis}

\subsection{Integrated spectrum}
\label{section_int_spectrum}

\begin{figure}
	\resizebox{\hsize}{!}{\includegraphics{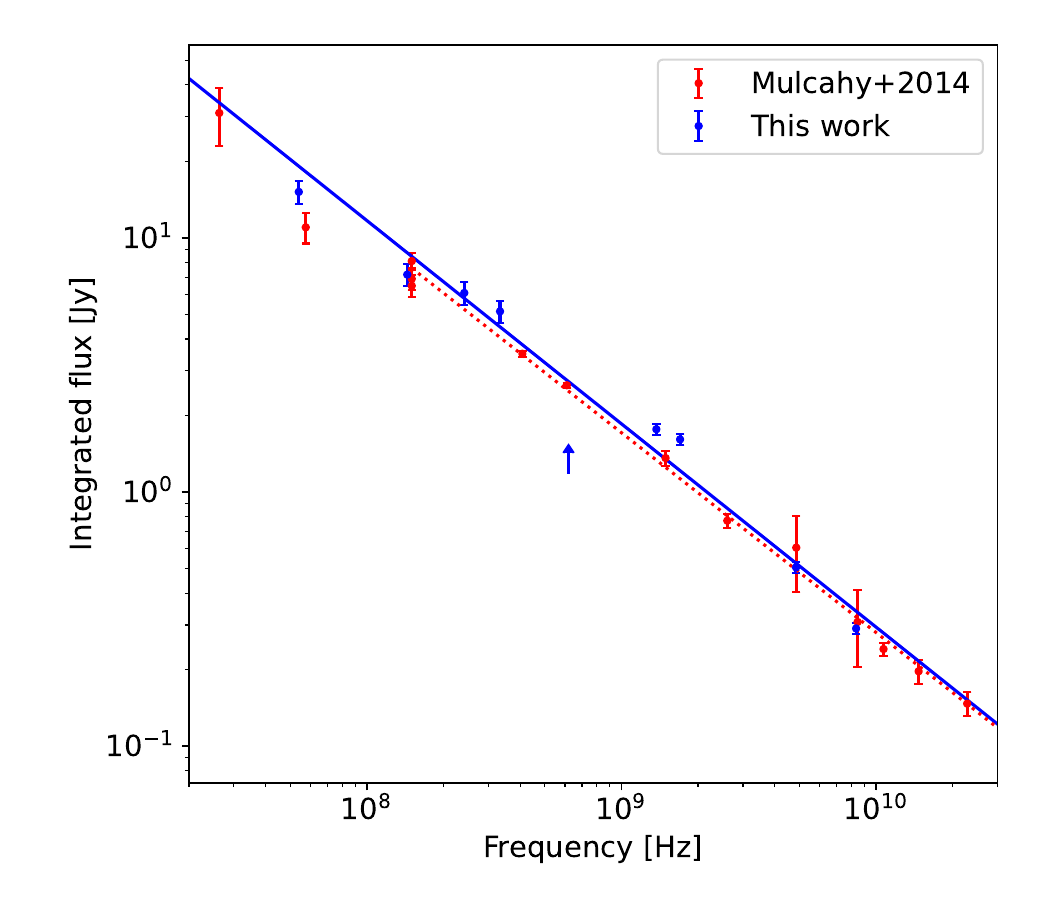}}
	\caption{Global radio continuum spectrum of M\,51. We show integrated flux densities calculated from our data and compare them with literature data compiled by \citet{2014A&A...568A..74M}. Power-law fits correspond to radio spectral indices of $-0.80\pm0.05$ for our data (blue data points and solid line) and $-0.79\pm0.02$ for the literature data (red data points and dotted line).  The flux density at 619\,MHz is only given as a lower limit and not included in the fit.}
	\label{plot_integrated}
\end{figure}

We start our analysis by calculating the integrated spectrum, which can help us to assess the robustness of the data. The spectrum is plotted and compared to the integrated spectrum from \citet{2014A&A...568A..74M} in Fig.~\ref{plot_integrated}. The integrated spectrum follows a power law with a spectral index of $\alpha=-0.80\pm0.05$ without any flattening at low frequencies. This is consistent with the spectral index of $-0.79\pm0.02$ calculated by \citet{2014A&A...568A..74M}. This test confirms that the integrated flux density of M51 in our maps matches previous observations. 

\subsection{Separation of thermal and synchrotron emission}
\label{section_thermal_separation}

In order to get a better handle on the contribution of the thermal emission and separate the thermal from the non-thermal emission we used H$\upalpha$ to estimate the free--free emission in M\,51. For this, we used a continuum-subtracted H$\upalpha$ map obtained with the Kitt Peak National Observatory 2.1-m telescope using the narrow-band H$\upalpha$-filter KP1563 \citep{2003PASP..115..928K}. The map was downloaded from the ancillary data at the SINGS webpage.\footnote{\url{http://irsa.ipac.caltech.edu/data/SPITZER/SINGS/}} The H\,$\upalpha$ image, shown in Fig.~\ref{fig:subtraction_appendix}, has an angular resolution of $1\farcs 35 \times 1\farcs 35$ and an rms noise of $\approx$$50\,\upmu\rm Jy\, beam^{-1}$. Because the H\,$\upalpha$ emission is easily absorbed by the dust, the observed H\,$\upalpha$ intensity $I_{\rm H\upalpha, obs}$ needs to be corrected for foreground and internal extinction. 

The foreground extinction is low in the direction of M\,51 \citep{2011ApJ...737..103S}; for that reason we neglected it. The internal extinction within M\,51 was corrected with a $24\,\upmu$m mid-infrared map to obtain an extinction-corrected H$\upalpha$ emission line flux \citep{2009ApJ...703.1672K}:
\begin{equation}
F_{\rm H\upalpha} = F_{\rm H\upalpha, obs} + 0.02\,\nu_{\rm 24\upmu m}\,I_{\rm 24\upmu m}.
\end{equation}
Here, $I_{\rm 24\upmu m}$ is the intensity at $24\,\upmu$m and is obtained from a {\it Spitzer} map observed as a part of the SIRTF Nearby Galaxies Survey \citep[SINGS;][]{2003PASP..115..928K}.  The 24\,$\upmu$m map has an angular resolution of $6\arcsec$. We convolved the  observed H$\upalpha$ map to $6\arcsec$ and aligned it to the the same coordinate system as that of the $24\,\upmu$m map.

The thermal contribution to the radio continuum emission at a given frequency $\nu$ can then be calculated using \citep{1997A&A...323..323D}
\begin{equation}
	\begin{aligned}
	\frac{S_{\rm th}(\nu)}{\textnormal{erg\,cm}^{-2}\,\textnormal{s}^{-1}\,\textnormal{Hz}^{-1}} = 1.14 \times 10^{-14} \left(\frac{\nu}{\textnormal{1000 MHz}} \right)^{-0.1} \\
    \times \left( \frac{T_{\rm e}}{10^4\,\textnormal{K}} \right)^{0.34} \left( \frac{F_{\textnormal{H}\upalpha}}{\textnormal{erg\,cm}^{-2}\,\textnormal{s}^{-1}} \right),
			\label{Equation_thermfromha}
	\end{aligned}
\end{equation}
where $T_{\rm e}$ is the thermal electron temperature.

\begin{figure}
	\resizebox{\hsize}{!}{\includegraphics{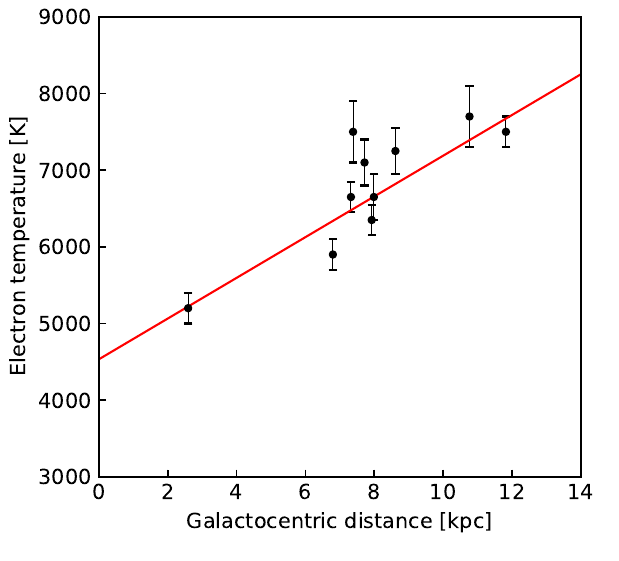}}
	\caption{Electron temperature $T_{\rm e}$ measured in H\,{\sc ii} regions as function of galactocentric distance $R_{\textnormal{gal}}$. Data taken from \citet{2004ApJ...615..228B}. The red line represents the best linear fit.}
	\label{plot_Te}
\end{figure}  
The electron temperature of the ionised gas phase is usually anti-correlated with the metallicity of the corresponding H\,{\sc ii} regions \citep{1979A&A....78..200A, 1979MNRAS.189...95P, 1981A&A....93..362S}. Since the metallicity usually increases towards the centre of galaxies, $T_{\rm e}$ decreases. In order to address the radial variation of $T_{\rm e}$, we used the catalogue of H\,{\sc ii} regions in M\,51 from \citet{2004ApJ...615..228B} and followed the linear fitting procedure of \citet{2014A&A...568A...4T} to calculate its radial profile. The  data with the fit are shown in Fig.~\ref{plot_Te}. The electron temperature $T_{\rm e}$ is found to follow the relation
\begin{equation}
	\frac{T_{\rm e}}{\rm K} = (266 \pm 47)\, \left( \frac{R_{\mathrm{gal}}}{\mathrm{kpc}} \right) + (4533 \pm 374),
	\label{Equation_Te}
\end{equation}
where $R_{\mathrm{gal}}$ is the galactocentric distance. We note here that the $T_{\rm e}$ measured in \citet{2004ApJ...615..228B} are derived from auroal lines, which are only detectable in low-metallicity H\,{\sc ii} regions. This could cause a bias towards higher temperatures in our profile, since temperature and gas metallicity anti-correlate, and so for higher metallicities one is not anymore able to detect auroral lines.  However, the overall radial dependence should still represent the average electron temperature in the H$\upalpha$ gas.

The extinction-corrected H$\upalpha$ map was then convolved to our common beam of $17 \farcs 5 \times 15\farcs 0$, and re-gridded to the same coordinate system as the radio maps.
For each pixel of a radio continuum map at a given frequency $\nu$, the thermal flux density was calculated using Eq.\,\eqref{Equation_thermfromha} and $T_{\rm e}$ from Eq.\,\eqref{Equation_Te}. The calculated thermal flux densities (Fig. \ref{fig:subtraction_appendix}) were subtracted from the total flux densities in our radio maps. The resulting non-thermal maps were used for the majority of the following analysis.

We found thermal contributions between 0.98\% and 4.8\% at 54\,MHz, and between 12\% and 14\% at 8350\,MHz.
Note that, for a relative error of $\varepsilon_{\rm th}$ in the estimated thermal fraction $f_{\rm th}$, the relative error in the synchrotron emission fraction ($\sigma_{ f_{\rm nth}}$; where $f_{\rm nth} = 1 - f_{\rm th}$) is given by $\sigma_{ f_{\rm nth}} = [1 - (1 \pm \varepsilon_{\rm th})\ f_{\rm th}]/[1 - f_{\rm th}]$ \citep{2017MNRAS.471..337B}.\footnote{Relative error of the thermal fraction $f_{\rm th}$ is defined as $\varepsilon_{\rm th} =  \sigma_{\rm f_{\rm th}}/f_{\rm th}$, where $\sigma_{f_{\rm th}}$ is the absolute error of $f_{\rm th}$.} That means an error of up to 20\,\% ($\varepsilon_{\rm th} = 0.2$) in a region with $f_{\rm th} = 0.15\,(0.05)$ will propagate to an error of about 5\,(2)\,\% in the estimated synchrotron emission. Adding this (in quadrature) to the flux uncertainty will increase it from 5(10)\,\% to 6.6(10.2)\,\% assuming an exaggerated error of 20\% on the thermal fraction. Because of this, the error on the thermal fraction will not significantly affect results presented in the rest of this paper.

\subsection{Spiral arm, inter-arm, and galaxy core regions}
\label{section_spiral_inter-arm}

Previous work \citep{2011MNRAS.412.2396F} has shown the radio continuum spectrum to be different depending on the location in the galaxy. In spiral arms the radio spectral index indicates a flat non-thermal spectrum, whereas the inter-arm regions have steeper spectra. Also, the radio continuum emission depends on the star-formation rate (SFR), and this radio--SFR relation is different in the arm and inter-arm regions. Hence, we now define arm- and inter-arm regions using the distribution of the atomic and molecular gas.

We decided to follow the approach of \citet{2009A&A...495..795H} using a combination of H\,{\sc i}- and CO-maps to generate a gas surface density map, which then represents the spiral arm structure tracing the density waves \citep{Colombo2014}. To this end, we used the integrated H\,{\sc i} map (moment 0) from \citet{2008AJ....136.2563W} and the integrated $\rm ^{12}CO$ 2-1 map from \citet{2007A&A...461..143S}. Both maps were  convolved to our common resolution of $17\farcs 5\times 15\farcs 0$. We calculate the H\,{\sc i} column density $N_{\textnormal{H\,{\sc i}}}$ using \citep{2017PASA...34...52M}:
\begin{equation}
 N_{\textnormal{H\,{\sc i}}} = 1.10\times 10^{24} \,\textrm{cm}^{-2} \left ( \frac{S_{\textnormal {H\,{\sc i}}}}{\rm Jy\,km\,s^{-1}}\right ) \left  (\frac{a\times b}{\rm arcsec^2} \right )^{-1},
\end{equation}
where $S_{\rm H\,{\textnormal{\sc i}}}$ is the velocity integrated H\,{\sc i} flux density, and $a$ and $b$ are the major and minor axis of the synthesised beam defined as full width at half maximum, respectively. 

\begin{figure}[t]
	\resizebox{\hsize}{!}{\includegraphics{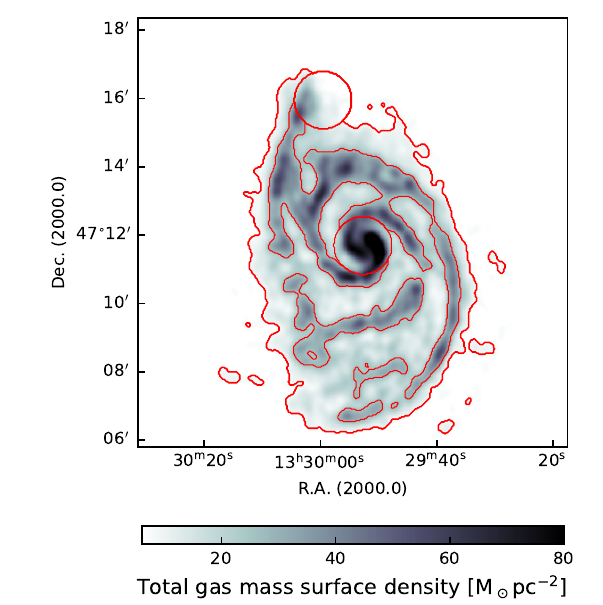}}
	\caption{Total gas mass surface density as  derived from a combination of atomic (H\,{\sc i}) and molecular ($\rm H_2$) gas maps. Contours at $8$ and $25\,\rm M_{\odot}\,pc^{-2}$ define the borders of the spiral arm and inter-arm regions, respectively. Circular apertures denote the core regions in, both, M\,51 and NGC\,5195. }
	\label{image_regions}
\end{figure}

In order to convert the CO map to an H$_2$ column density map we use \citep{2007A&A...461..143S}:
\begin{equation}
	N_{\rm H_2} = 0.25\  \frac{X_{\rm MW}}{0.8} \frac{T_{\rm CO}}{\rm K\,km\,s^{-1}},
\end{equation}
where $X_{\rm MW}=2.3\times 10^{20}\,\rm cm^{-2}\, (K\,km\,s^{-1})^{-1}$ is the CO-to-H2 conversion factor as determined for the Milky Way \citep{2007A&A...461..143S} and $T_{\rm CO}$ is the velocity integrated CO intensity given as main beam antenna temperature. The factor $0.8$ reflects the assumed 2-1/1-0 CO intensity ratio.

Both atomic and molecular gas maps were first converted to their corresponding mass surface densities. To convert from the observed integrated intensities to the total gas surface densities, we use the relation $\Sigma_{\rm gas}=1.36\,(\Sigma_{\textnormal {H\,{\sc i}}}+\Sigma_{\rm H2})$ taking into account the mass contribution from He. The distribution of $\Sigma_{\rm gas}$ is presented in Fig.\,\ref{image_regions}. As in \citet{2009A&A...495..795H}, we use total mass surface density threshold in order to define spiral arm regions. These regions were chosen to be similar to those in interferometric CO maps, in H$\upalpha$ maps, and in 1400\,MHz radio continuum maps. All pixels with $\Sigma_{\rm gas}>25\,\rm M_{\odot}\,pc^{-2}$ were attributed to the spiral arm regions and pixels with  $8\,\rm M_{\odot}\,pc^{-2}\leq \Sigma_{\rm gas}\leq 25\,\rm M_{\odot}\,pc^{-2}$ were attributed to the inter-arm regions. We defined two additional regions in the cores of M\,51 and NGC\,5195 with a radius of 25\arcsec. The resulting map and the defined regions are shown in Fig.~\ref{image_regions}.

\subsection{Non-thermal radio spectral index and curvature}
\label{section_SI_SC}

\begin{figure*}[t]
	\resizebox{\hsize}{!}{\includegraphics{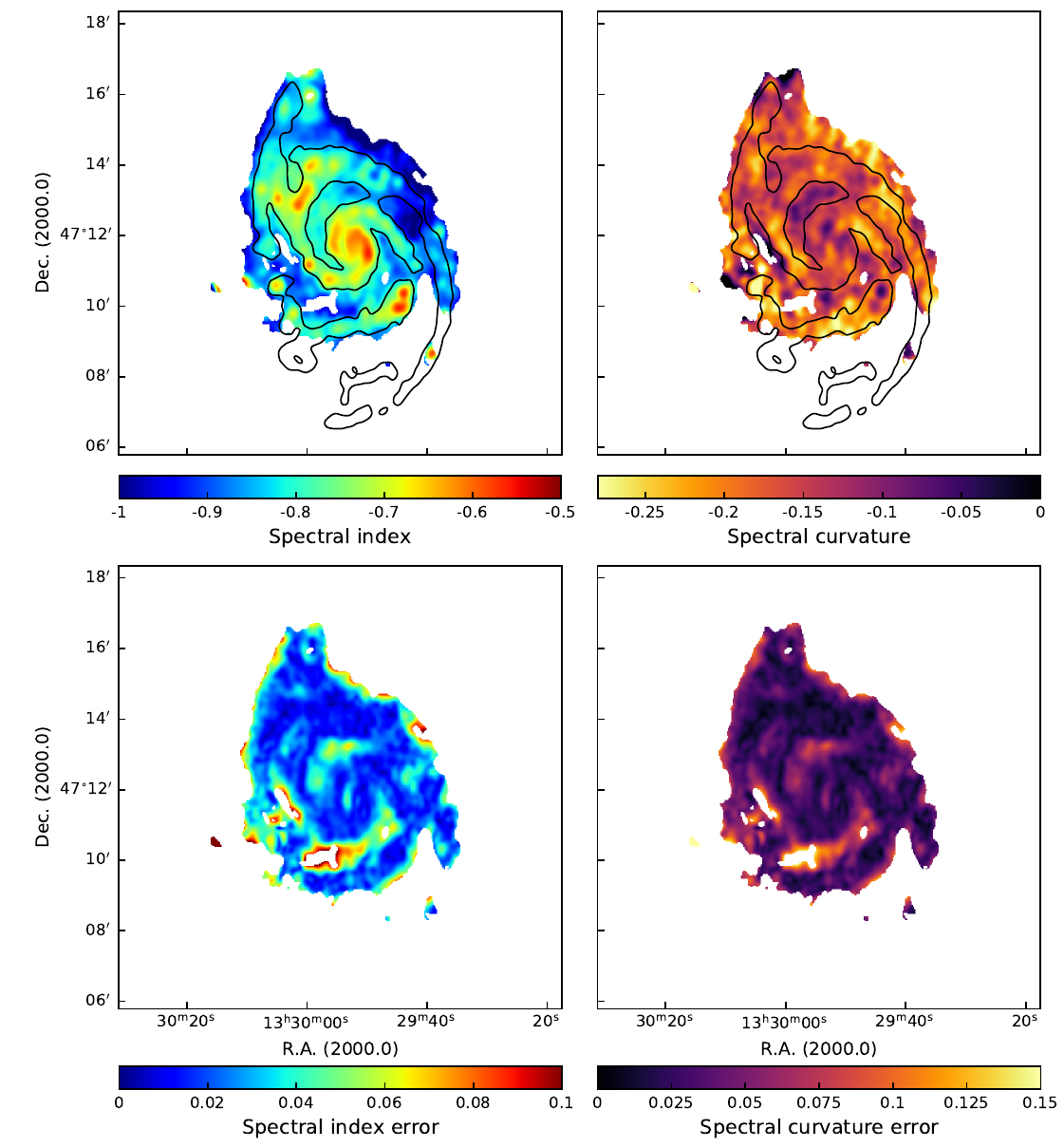}}
	\caption{Non-thermal radio spectral index and curvature at 1\,GHz. Top panels: Best-fit radio spectral index $\alpha_0$ (left panel) and spectral curvature (right panel). Bottom panels show the corresponding uncertainty maps of the radio spectral index (left panel) and curvature (right panel). The black contours represent a total gas mass surface density 25\,M$_\odot\, \rm pc^{-2}$.}
	\label{image_SI_SC}
\end{figure*}

In the next step, we explore the observed non-thermal radio continuum spectra in different regions of the galaxy on a point-by-point basis. We fit a model spectrum consisting of a power law and a curvature component using a polynomial equation in logarithmic space \citep{2013ApJS..204...19P, 2017ApJS..230....7P}: 
\begin{equation}
	\log S_\nu = \log S_0 + \alpha_0 \log(\nu/\nu_0) + \beta\,\left[\log (\nu/\nu_0)\right]^2,
	\label{equation_spectrum}
\end{equation}
where $S_\nu$ is the flux density at a given frequency, $\nu$ in GHz, $S_0$ is the flux density at the normalization frequency $\nu_0=1$\,GHz, $\alpha_0$ is the non-thermal radio spectral index, 
$\beta$ the non-thermal radio spectral curvature. Note that $\beta<0$ corresponds to a concave spectrum. The non-thermal radio spectral index $\alpha_0$ needs to be defined at a single reference frequency, because the spectral slope is frequency dependant in case of a curved spectrum. 
We start with this purely phenomenological model in order to explore the general shape of the spectrum without bias toward a specific physical mechanism. We explore the underlying physics responsible for the shape of the spectrum in Sect. \ref{section_turnover}. 

\begin{figure}
	\resizebox{\hsize}{!}{\includegraphics{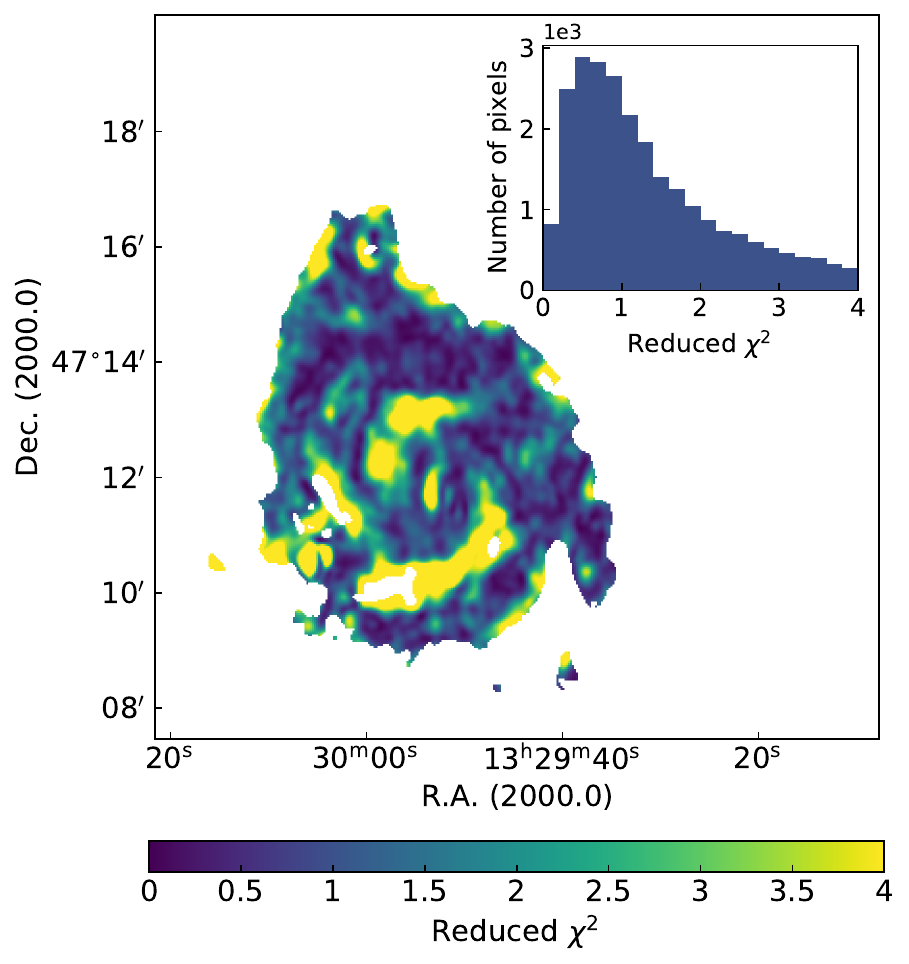}}
	\caption{Map of the reduced $\chi^2$ of the fit presented in Fig. \ref{image_SI_SC}. The inset on the top right shows the histogram of the reduced $\chi^2$ values. }
	\label{image_chi2}
\end{figure}

For each point in the maps we fit this model to eight data points at frequencies listed in table \ref{table_data} (excluding 619\,MHz). For simplicity, we ignore higher-order terms and more sophisticated modeling of the spectra needs to be deferred to future work. The data are fitted with Levenberg--Marquardt least-squares algorithm. The resulting 
maps of best-fit $\alpha_0$ and $\beta$, and their corresponding error maps are shown in Fig.~\ref{image_SI_SC}. The reduced $\chi^2$ of the fit is shown in Fig~\ref{image_chi2}. The resulting parameter values are restricted to the frequency range between 54 and 8350\,MHz and cannot be extrapolated towards arbitrary high or low frequencies.

The non-thermal radio spectral index at 1\,GHz has values between $-1.0$ and $-0.5$. The flatter radio spectra ($\alpha_0$ from $-0.7$ to $-0.5$) are mostly found along the spiral arms where the star-forming regions are located. 
The inter-arm regions have radio spectral indices between $-1.0$ and $-0.8$. The tidal bridge region between M\,51 and NGC\,5195 (located north of M51) is characterized by a similarly steep radio spectrum, whereas the spectrum flattens again towards the companion with spectral indices of $\alpha_0\approx -0.7$. Errors lie usually below $0.04$ and only rise to $0.10$ in regions of lower signal-to-noise ratios. Our radio spectral index map is very similar to the 1400--4850\,MHz map by \citet{2011MNRAS.412.2396F}, who used a $1400$\,MHz map from the VLA instead of the WSRT map (and the same $4850$\,MHz map we use). Our radio spectral index is defined at a lower frequency of 1\,GHz, and so deviations of $0.1$--$0.2$ become apparent in regions where the absolute value of the spectral curvature is high. 

The spectral curvature is negative with values between $-0.3$ and $0.0$. This means the radio spectrum is concave, implying a suppression of radio emission at either low or high frequencies, or a combination of both. The suppression of radio emission at low frequencies can happen be due to free--free absorption, ionisation losses or synchrotron self absorption and at high frequency can happen due to CR radiation losses \cite{2011hea..book.....L}. We further explore which process is responsible for the observed curvature in Sect. \ref{section_turnover}. Regions with flatter radio spectra tend to show stronger curvatures ($\beta\leq-0.2$). Errors for the curvature are usually smaller than $0.06$; only in regions with lower signal-to-noise ratios they can rise to $0.14$.
The reduced $\chi^2$ is smaller than one in most of the galaxy, the value is higher in the areas where the error on the fit parameters is also high. The areas where the model doesn't fit the data so well also have a somewhat lower curvature. We assume that a better model in those areas would be just a power law. 

\subsection{Low- and high-frequency non-thermal radio spectral indices}
\label{section_low_high}

\begin{figure*}
	\resizebox{\hsize}{!}{\includegraphics{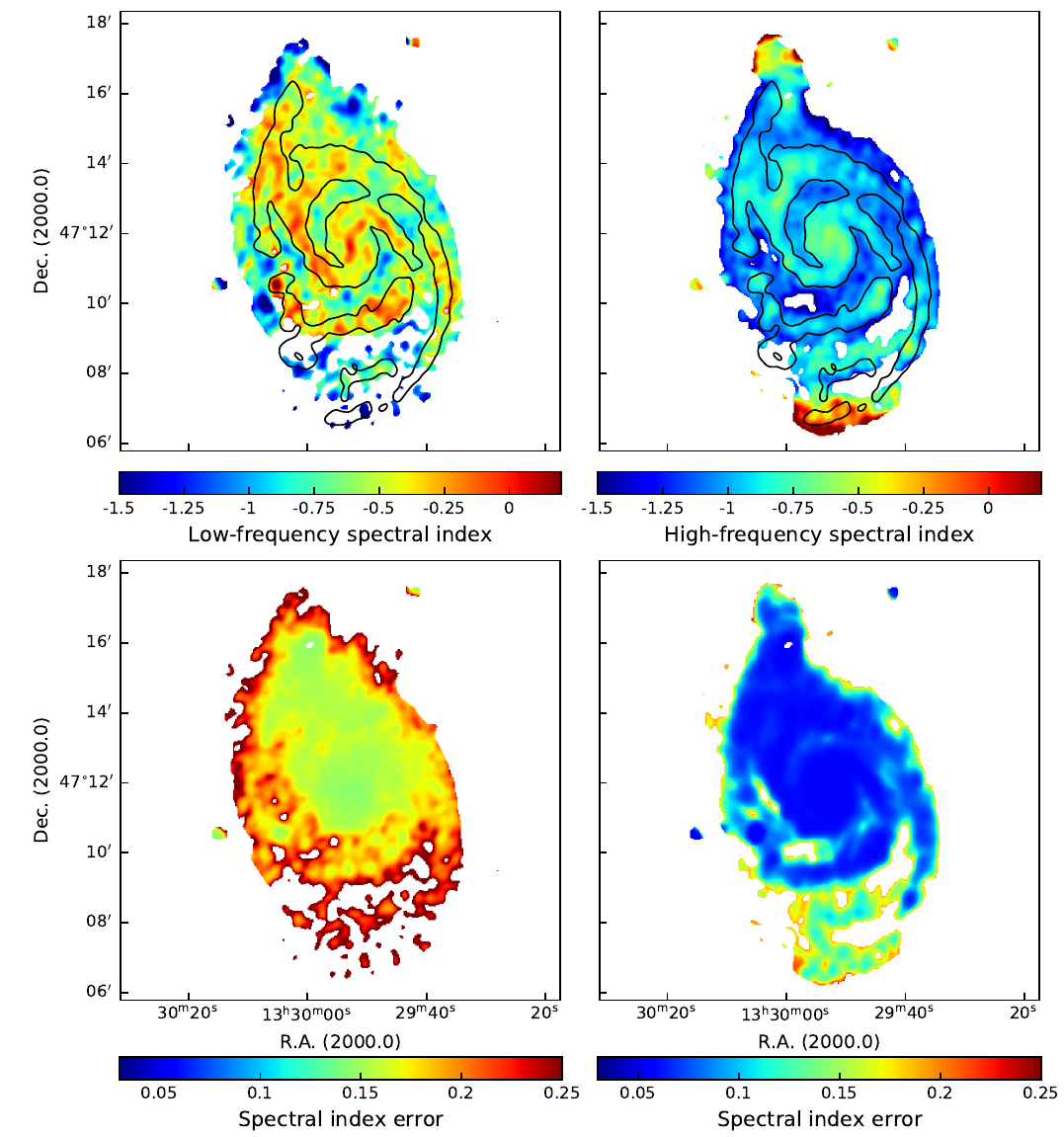}}
 	\caption{Two point non-thermal radio spectral indices at low ($\alpha_{\rm low}$; 54--144\,MHz) and high ($\alpha_{\rm high}$; 1370--4850\,MHz) frequencies. The top panels show the spectral indices $\alpha_{\rm low}$ (left panel) and at $\alpha_{\rm high}$ (right panel). Bottom panels show the corresponding uncertainty maps. The black contours represent a total gas mass surface density 25\,M$_\odot\, \rm pc^{-2}$.}
	\label{image_SI}
\end{figure*}

We now study the non-thermal spectral index both at low and high frequencies. We compare the two-point low-frequency radio spectral index $\alpha_{\rm low}$ between 54 and 144\,MHz with the two-point high-frequency spectral index $\alpha_{\rm high}$ between 1370 and 4850\,MHz. The two-point spectral indices were calculated in a standard way by taking the logarithmic ratio of the flux densities in two maps over the logarithmic ratio of frequencies. The errors in the spectral index maps are calculated by propagating the errors from the individual maps which are a combination of the flux error (listed in Table \ref{table_data}) and the rms noise measured away from the source. The two-point spectral index maps are shown in Fig.~\ref{image_SI}.
The reason for not selecting the two highest frequencies is the smaller extent of the galaxy at 8350\,MHz, which would prevent us from studying the radio spectral index of the diffuse emission further in the outskirts of M\,51. Additionally, skipping the 8350\,MHz map gives a higher ratio between frequencies which results in a more reliable spectra index map.

On a cursory look, the maps of $\alpha_{\rm low}$ and $\alpha_{\rm high}$ have a similar morphology, but $\alpha_{\rm low}$ is systematically higher than $\alpha_{\rm high}$. Values for $\alpha_{\rm low}$ reach from $-$0.7 to 0.2 with errors in the range of $0.17$--$0.25$. In contrast, values for $\alpha_{\rm high}$ reach from $-1.3$ to $-0.6$ with errors in the range of  $0.05$--$0.15$. Higher errors for the low-frequency spectral index are expected because of the higher calibration error and noise in those maps. The errors are also higher at the edges of the maps because of a lower signal-to-noise ratio. The higher values of $\alpha_{\rm high}\approx-0.7$ correlate well with the spiral arm structure of M\,51, which has already been shown by \citet{2011MNRAS.412.2396F}. The values of $\alpha_{\rm low}$ are of similar apparent distribution, although the spiral arms are more diffuse and have an inverted radio spectrum as indicated by radio spectral indices with values of up to $\sim$$0.2$. Because of the cosmic ray radiation losses of the CR electrons, we expect a steepening at the edges of the galaxy. 
This is exactly what we see in the $\alpha_{\rm low}$ map even if it might not be completely reliable because of the low signal-to-noise in these regions. We attribute the strong flattening of the radio spectrum at higher frequencies towards the extreme north and south as visible in the map of $\alpha_{\rm high}$ as spurious. They are likely due to the limitations of combining the interferometric and single-dish data \citep{Cotton2017}.  These data do not affect our results as they are outside of our region of interest that we consider for our analysis (Sect.\,\ref{section_data}).

As the results of fitting for spectral index and curvature already suggested (Fig. \ref{image_SI_SC}), we see an overall flattening of the spectral index towards lower frequencies for the whole inner disc of M\,51. In the following, we discuss the possible effects causing these flat and partially inverted spectra.

\section{Spectral index flattening and low-frequency turnovers}
\label{section_turnover}

Changes of spectral indices over frequency may be due to a number of reasons. 
In the following we investigate the plausible mechanisms that can altering the radio continuum spectrum from a power-law to a concave spectrum. This can be done either by  changing the cosmic-ray electron spectrum or by directly absorbing low-frequency radio emission. Previous work has shown \citep{2023A&A...672A..21H, 2024A&A...682A..83H} that the diffusion coefficient of GeV CR electrons is energy-independent, at least over the relevant range in energies. Hence, we do not expect diffusion to change the spectrum. Also, advection does not change the spectrum. Hence, we now consider cosmic-ray radiation losses, synchrotron self-absorption, thermal free--free absorption, and cosmic-ray ionisation losses.

\subsection{Cosmic-ray radiation losses}
\label{ss:cr_radiation_losses}

In regions of low gas densities, cosmic-ray electrons lose their energy  due to synchrotron and inverse Compton radiation losses, causing their spectra to age, i.e. steepen.
 
In particular, spectral ageing predominantly affects the high-frequency radio spectral index (Sect.\,\ref{section_low_high}). We find that near star-forming regions the values of $\Delta\alpha= \alpha_{\rm low} - \alpha_{\rm high}$ are lower (between 0.2 and 0.5) compared to the inter-arm regions, where they are between 0.8 and 1.1.
A proper modeling of CR ageing is not possible given the data but we can make a few qualitative observations. Fig.~\ref{image_SI_SC} clearly shows that the curvature is high, and the spectral index is lower in regions with low star-formation, i.e., in the inter-arm regions. Inside the spiral arms, the synchrotron emission at low frequencies is likely to be a mix of recently accelerated CR electrons and those from previous, nearby sites of star formation which fall within your resolution element. However, this is not true in the inter-arm regions where star formation is low. We have a resolution of aproximately 16", i.e. 650 kpc. Assuming that the bulk of the CR electrons are produced in the arms, then CR electrons with energies lower than about 1.5 GeV will be able to diffuse to scales larger than 650 kpc (assuming $B=10\,\mu$G, and $D=2\cdot 10^{28}$ cm$^2$/s). This corresponds to CR electrons emitting at critical frequencies of 600-700 MHz. Hence, CR electrons emitting above 1 GHz, will lose energy before they get the chance to mix, especially in the inter-arm regions. Note, even in low star formation regions, the local CR electron injection timescale is much longer than the aging timescale. This is quantitatively shown in Fig.~3 of \cite{2015MNRAS.449.3879B} in order to explain steep spectra in low-density regions.  This argument is also supported by spatially resolved radio-FIR relation \citep[see Fig. 5 and Section 4 of][]{2012ApJ...756..141B}. 

\subsection{Synchrotron self-absorption}

Synchrotron self-absorption is known to play a key role in the flattening of spectra in the inner cores of Active Galactic Nuclei (AGN). We use the equation from \citet{2013MNRAS.431.3003L} to determine the turnover frequency $\nu_{\rm SSA}$ for integrated spectra of galaxies

\begin{equation}
\nu_{\rm SSA} = 2.4\,{\rm MHz} \left( \frac{\Sigma_{\rm SFR}} {50\,\text{M}_\odot \text{yr}^{-1} \text{kpc}^{-2}} \right)^{0.39} ,
\end{equation}
where $\Sigma_{\rm SFR}$ is the star-formation rate surface density. With a value of $\Sigma_{\rm SFR} = 0.9$ from \citet{2013ApJ...770...85G}, we calculate a value of $\nu_{\rm SSA}=0.5\,$MHz. This value is by more than two orders of magnitude lower than our lowest observing frequency. Therefore, synchrotron self-absorption cannot be a dominant effect causing the flattening.

Similarly, diffuse large-scale structures in nearby galaxies are not very likely to be affected by synchrotron self-absorption due to the needed high magnetic field strength. We can calculate this magnetic field strength $B_{\rm SSA}$ for a given frequency $\nu$ in GHz and brightness temperature $T_{\rm b}$ in K using the standard equation:

\begin{equation}
	B_{\rm SSA} \approx 1.4\cdot 10^{21} \nu T_{\rm b}^{-2} .
\end{equation}
$T_{\rm b}$ is defined as

\begin{equation}
	T_{\rm b} = 1.222\cdot 10^3 \frac{I}{\nu^2 \theta_{\rm maj} \theta_{\rm min}} \ \mathrm{K} ,
\end{equation}
with the brightness $I$ in Jy/beam and $\theta_{\rm maj}$ and  $\theta_{\rm min}$ the synthesised major and minor beam half-power beam widths in arcseconds, respectively. Inserting our beam size of $17.5\arcsec\times 15\arcsec$ and a brightness of 150\,mJy/beam at 54\,MHz as a maximum we end up with a magnetic field of $1.3\times10^{15}$\,G. From equipartition calculations we know that the typical magnetic field strength in nearby galaxies is $9\,\upmu$G \citep{2005mpge.conf..193B} with maximum values for extreme starburst cases like M\,82 of $100\,\upmu$G \citep{2013A&A...555A..23A}. \citet{2013MNRAS.431.3003L} discussed the possible turnover frequency for extreme starburst cases, which would be around $\nu_{\rm SSA}=2.4$ MHz. These results show that synchrotron self-absorption cannot play an important role in the flattening of the spectra at low frequencies in M\,51. 

\begin{table*}[t]
	\caption{Fitted parameters for the spectra of the spiral arm (SA), inter-arm (IA) and core (N5194c, N5195c) regions. Columns (2)--(4) are for free--free absorption and Columns (5)--(7) are for the curved spectral model, both are fitted to the non-thermal radio continuum spectra.}
	\label{table_spectra}
	\centering
	\newcolumntype{0}{>{\centering\arraybackslash} m{1.5cm} }
	\newcolumntype{1}{>{\centering\arraybackslash} m{2.2cm} }
	\newcolumntype{2}{>{\centering\arraybackslash} m{2.2cm} }
	\newcolumntype{3}{>{\centering\arraybackslash} m{2.2cm} }
	\newcolumntype{4}{>{\centering\arraybackslash} m{2.2cm} }
	\newcolumntype{5}{>{\centering\arraybackslash} m{2.2cm} }
	\newcolumntype{6}{>{\centering\arraybackslash} m{2.2cm} }
	\renewcommand{\arraystretch}{1.5}
	\begin{tabular}{@{} 0 1 2 3 4 5 6 @{}}
		\toprule
		\toprule
		Region & $S_{0}$ [Jy] & $\eta$ & EM [pc cm$^{-6}$]& $S_{0}$ [Jy] & $\alpha_0$ & $\beta$ \\
		\midrule
        SA$_{\rm syn}$ & $0.630\pm 0.036$ & $-0.839\pm 0.039$ & $2800\pm 990$ & $0.697\pm 0.013$ & $-0.777\pm 0.0089$ & $-0.176\pm 0.013$ \\
        IA$_{\rm syn}$ & $0.459\pm 0.026$ & $-0.914\pm 0.041$ & $2000\pm 790$ & $0.507\pm 0.026$ & $-0.881\pm 0.024$ & $-0.147\pm 0.034$ \\
        N5194c$_{\rm syn}$ & $0.293\pm 0.014$ & $-0.746\pm 0.032$ & $2710\pm 980$ & $0.3119\pm 0.0071$ & $-0.674\pm 0.012$ & $-0.144\pm 0.016$ \\
        N5195c$_{\rm syn}$ & $0.0917\pm 0.0036$ & $-0.865\pm 0.027$ & $2040\pm 570$ & $0.09924\pm 0.0021$ & $-0.821\pm 0.0097$ & $-0.132\pm 0.014$ \\
		\bottomrule
	\end{tabular}
\end{table*}

\subsection{Thermal free--free absorption}

Free--free absorption is an effect caused by an ionised medium, which absorbs radio waves while they propagate through it. The amount of absorption depends on the thermal electron number density $n_e$, the path length, and the frequency, such that the optical depth increases with decreasing frequency. Free--free absorption has been observed at frequencies below 500--1000\,MHz in the core regions \citep{2013A&A...555A..23A,2016A&A...593A..86V} and individual star-forming regions \citep{1997MNRAS.291..517W,2015A&A...574A.114V,2017MNRAS.471..337B} of starburst galaxies, and also in the Milky Way \citep{1958MNRAS.118..591R,2004MNRAS.349L..25R}. Observations at very low radio frequencies also showed a turnover for the global spectrum of the Milky Way at about 3\,MHz \citep{1973ApJ...180..359B}.

In the following we investigate if free--free absorption can explain the flattening of the radio spectrum towards lower frequencies. Following \citet{1997MNRAS.291..517W}, the spectrum in case of free--free absorption of sources inside an ionised medium is
\begin{equation}
	S_\nu = S_0 \nu_0^{\eta} e^{-\uptau_{\rm ff}} ,
	\label{equation_thermabs}
\end{equation}
with $\eta$ the radio spectral index of the optically thin medium, and $\uptau_{\rm ff}$ the free--free emission optical depth. Taking the expression for the free--free Gaunt factor at radio wavelengths from \cite{2005ism..book.....L}, the optical depth is given as
\begin{equation}
 \uptau_{\rm ff} = {8.2 \times 10^{-2}\, \nu_0^{-2.1}\, {T_{\rm e}^{-1.35}}\, {\rm EM}},
	\label{equation_opacity}
\end{equation}
with the emission measure (EM) in units of cm$^{-6}$\,pc and the electron temperature $T_{\rm e}$ in K. The EM is defined as 
\begin{equation}
	\left(\frac{\mathrm{EM}}{\rm pc\,cm^{-6}}\right) = \int_{0}^{s_0} \left( \frac{n_e}{\rm cm^{-3}} \right)^2\ \left( \frac{{\rm d}s}{\rm pc} \right),
\end{equation}
where $n_e$ is the thermal free electron density, and $s$ is the distance along line-of-sight between the source at $s_0$ and observer at 0. The absorption model (Eq.\,\ref{equation_thermabs}) was used to fit the spectrum of M\,51. We then derived the EM using Eq.\,\eqref{equation_opacity}, where we inserted the electron temperature as expressed by the galactocentric distance (Eq.\,\ref{Equation_Te}). 

\begin{figure}
	\resizebox{\hsize}{!}{\includegraphics{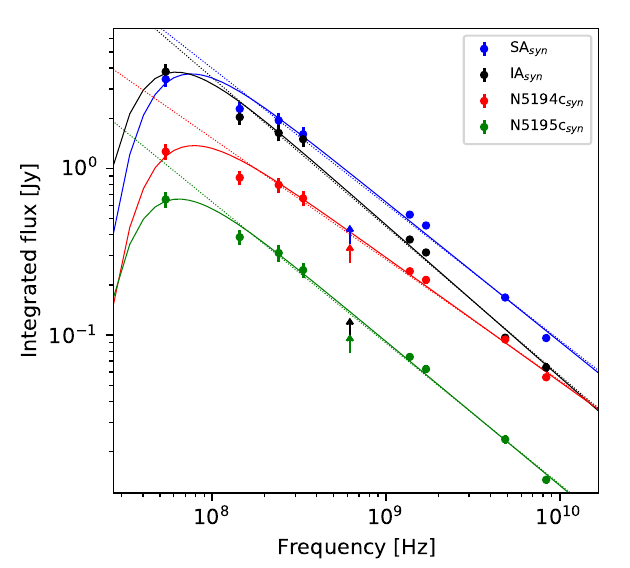}}
	\caption{Non-thermal radio continuum spectra for the spiral arm (SA), inter-arm (IA), and core regions (N5194c and N5195c). The flux densities at 619\,MHz are only given as lower limits and are not included in the analysis. Solid lines are the best-fitting free--free absorption models (Eq. \ref{equation_thermabs}). Dotted lines show the power-law fits for frequencies $\geq$144\,MHz.}
	\label{plot_SEDs}
\end{figure}

\subsubsection{Local radio continuum spectra}

We first investigate the effect of free--free absorption on our radio continuum spectra analysing the integrated flux densities within the four regions of the arm, inter-arm, and two cores of, both, M\,51 and NGC\,5195 (Sect.\,\ref{section_spiral_inter-arm}). The thermal emission which we have previously subtracted is also affected by free--free absorption.
However, we can ignore the absorption of the thermal radio continuum as free--free absorption is only significant at low frequencies where the thermal fraction is very low. Similarly, at high frequencies free--free absorption is negligible. We can therefore justify using the non-thermal maps as produced above for the fitting and neglect the effect of free--free absorption on the thermal emission. 
For comparison, we also fit the polynomial model (Eq.\,\ref{equation_spectrum}) to the non-thermal radio continuum spectra (after thermal subtraction) in the four regions. The flux density error was estimated by assuming a statistical error caused by the map noise $\sigma_{\rm rms}$ as well as a relative flux uncertainty $\varepsilon_\nu$ due to calibration uncertainty (see Table~\ref{table_data}). The error of the flux-density measurements at each frequency is calculated using the following expression \citep{2022A&A...664A..83H}:
\begin{equation}
	\sigma_{S_\nu} = \sqrt{\bigg(\sigma_{\rm rms}\sqrt{N_{\rm beams}}\bigg)^2 + \big(\varepsilon_\nu\ S_\nu\big)^2},
\end{equation}
where $S_\nu$ is the flux density and $N_{\rm beams}$ is the number of beams in the integration region.

Figure \ref{plot_SEDs} shows the measured spectra within the regions together with the best-fitting absorption models. We show additional point spectra in Appendix \ref{a:point}. The flattening of the spectra at low frequencies is highlighted by comparing the spectrum to a power law. The resulting parameters are listed in Table \ref{table_spectra}. The radio spectral indices $\eta$ lie between $-0.91$ and $-0.75$. For the polynominal fits, we find non-thermal radio spectral indices at 1\,GHz between $-0.88$ and $-0.67$.

We find that the EM has values between 2000 and 2800 cm$^{-6}$\,pc. They mostly show the same tendency as the absolute values of the spectral curvature, as the higher values of EM are seen in the spiral arm regions, and the lower values in the inter-arm regions. However, the values of EM do not show a statistically significant difference because of high uncertainties.

\begin{figure*}[ht]
	\resizebox{\hsize}{!}{\includegraphics{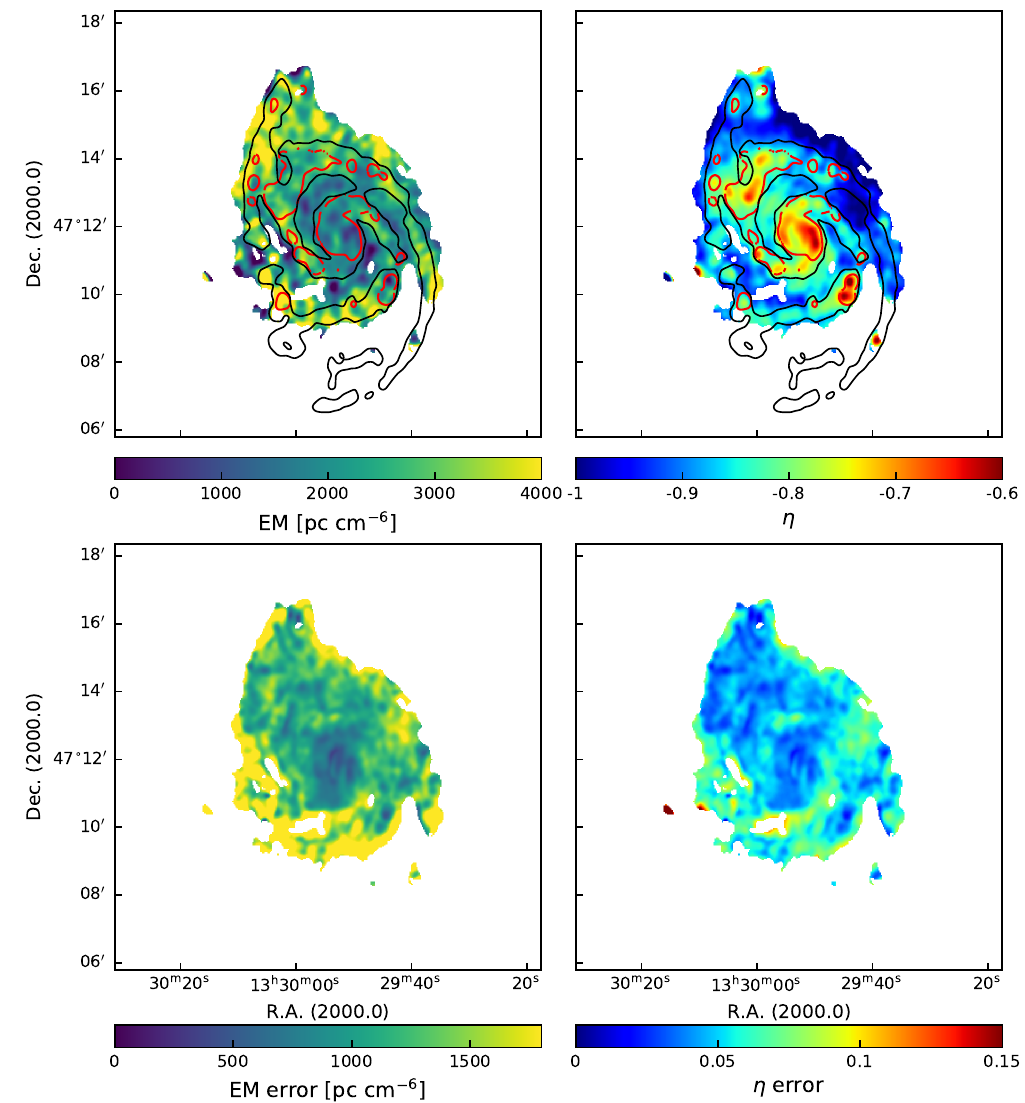}}
	\caption{Thermal free--free absorption: Shown are the results from fitting the thermal absorption model to the non-thermal radio continuum emission at eight frequencies between 45 and 8350 MHz (Table \ref{table_data}). The left column shows the resulting EM map (top panel) and its uncertainty map (bottom panel). The right column shows the radio spectral index of the optically thin medium $\eta$ (top panel) and its uncertainty map (bottom panel). The red contours represent the H$\upalpha$ flux value of $2.6\cdot 10^{-13}\,\mathrm{ erg\,s^{-1}\,cm^{-2}\,\AA^{-1}\,beam^{-1}}$ from the Metal-THINGS survey. The black contours represent a total gas mass surface density 25\,M$_\odot\, \mathrm{ pc^{-2}}$.}
	\label{image_thermabs}
\end{figure*}

\subsubsection{Emission Measure and radio spectral index}

We now investigate the free--free absorption on a point-by-point basis in order to determine the spatial distribution of the EM. For each point in the radio maps we fit the free--free absorption model (Eq. \ref{equation_thermabs}) to eight data points at frequencies listed in table \ref{table_data} (excluding 619\,MHz). We assume the electron temperature from Eq. \ref{Equation_Te}. The parameters of the model which we fit are the spectral index $\eta$ and the EM. We present the resulting maps in Fig.\,\ref{image_thermabs}. In general, regions with higher gas densities (the spiral arms) show flatter spectral indices and slightly higher EM. While the steeper spectral index in the inter-arm regions can be explained by radiation losses of the cosmic-ray electrons during propagation away from their places of origin in the star-forming regions, the significant flattening in the spiral arms and in the cores of NGC\,5194 and 5195 is difficult to explain without thermal absorption effects. The spectra presented in Fig.\,\ref{plot_SEDs} show that the 54\,MHz data points have much lower flux densities than what would be expected if the emission were to follow a power law. This is an indication of a flattening at low frequencies which is, at least in part, caused by free--free absorption.

We used the H\,$\upalpha$ flux density map obtained using integral field unit spectroscopy to calculate and independent estimate of EM. We first convolved the data to the same beam as the radio maps. To calculate the EM, we used the equation \citep{1992FCPh...15..143D}:
\begin{equation}
\mathrm{EM}= 5 \cdot 10^{17} F_{H\upalpha}/\Omega \ \mathrm{cm^{-6}pc,}
\end{equation}
where $F_{H\upalpha}$ is the measured flux in erg cm$^{-2}$ s$^{-1}$ and $\Omega=1.133\cdot 17 \farcs 5 \times 15\farcs 0$ is the collecting area in $\Box"$  calculated for our beam.

\begin{figure}
	\resizebox{\hsize}{!}{\includegraphics{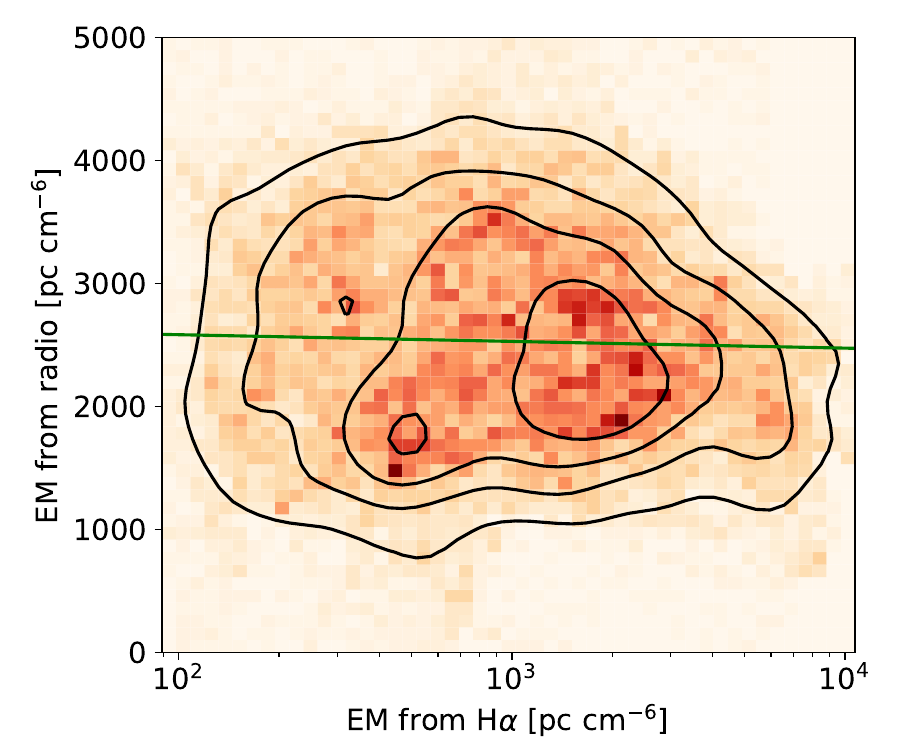}}
	\caption{Comparison of the EM from H$\upalpha$ flux density to the EM obtained by fitting the free--free absorption model to the radio data. Colours and contours represent the data point density. The green line shows the best-fitting, albeit insignificant, correlation. 
 }
	\label{plot_Ha_EM}
\end{figure}

We compare the EM from H$\upalpha$ flux density to the EM obtained by fitting the free--free absorption model to the radio data by plotting them against each other in Fig.~\ref{plot_Ha_EM}. Contrary to our expectation, we see no correlation with a Spearman rank correlation coefficient of $\rho_{\rm s}=-0.04$. This leads us to believe free--free absorption might not be the main mechanism responsible for the low-frequency flattening that we observe (see Sect. \ref{section_ion_losses}).  An alternative reason could be that H\,$\upalpha$ directly traces the clumpy ionised gas, while EM estimated from the radio spectrum also has contribution from the diffuse gas. The latter is known as the diffuse ionised gas that fills the space between the H\,{\sc ii} regions \citep{2009RvMP...81..969H}. 
The radio free--free emission also originates from such a medium. The lack of correlation, or putative anti-correlation, may be hence explained by the very different volume filling factors.

\subsection{Ionisation losses}
\label{section_ion_losses}

\begin{figure}
	\resizebox{\hsize}{!}{\includegraphics{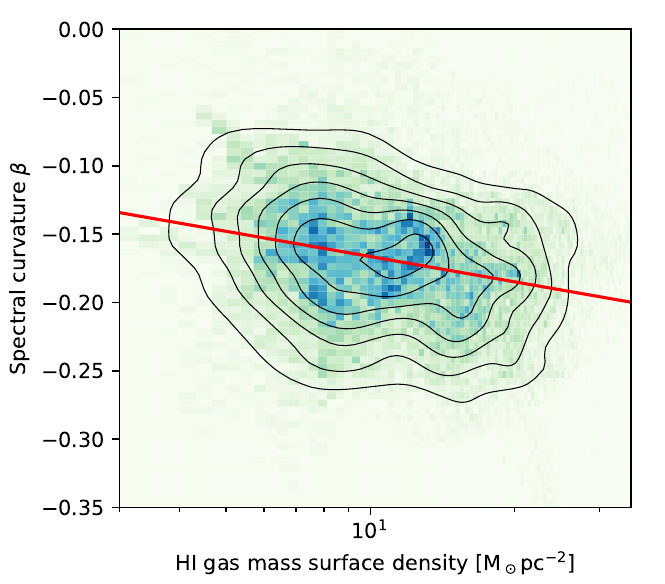}}
	\caption{Non-thermal radio spectral curvature  as  function of the H\,{\sc i} gas mass surface density. Colours and contours show the data point density. The red line shows the best-fitting correlation.
 }
	\label{plot_HI_EM}
\end{figure}

We do, however, find a correlation between the spectral curvature and the H\,{\sc i} gas mass surface density (Fig. \ref{plot_HI_EM}). This leads us to explore ionisation losses as an alternative mechanism to explain the low frequency flattening because it dominates in the neutral gas. 
Cosmic-ray electrons lose energy due to the ionisation of atomic and molecular hydrogen. The ionisation loss rate is directly proportional to the number density of neutral atoms and molecules $n$, as can be seen from the following expression in the case of neutral hydrogen \citep{2011hea..book.....L}:
\begin{equation}
	-\bigg( \frac{\mathrm{d}E}{\mathrm{d}t} \bigg)_{\rm ion} = 7.64 \times 10^{-15} n\ (3\,\mathrm{ln}\ \gamma +19.8)\ \rm eV\,s^{-1} ,
\end{equation}
where $E$ is the electron energy, and $\gamma$ is the Lorentz factor.
The dependence on energy is only logarithmic so it can be approximated as constant throughout the spectrum. At lower energies a larger fraction of energy is lost due to ionisation and therefore the effect is stronger. Note that the fraction of energy lost is higher at lower CR energies because the total energy is lower while the lost energy is approximately constant. This means that the cosmic-ray electron number density (per energy bin) does decrease as function of time and this effect is more pronounced at lower energies. The spectral index flattening due to ionisation losses should be $\Delta\alpha \leq 0.5$ in comparison to the injection spectrum \citep{2015MNRAS.449.3879B}.

In Fig.\,\ref{plot_HI_EM} we show the non-thermal spectral curvature as function of H\,{\sc i} gas mass surface density. We find a weak negative correlation ($\rho_{\rm s}=-0.21$). This can be explained if cosmic-ray ionisation losses are causing the low frequency flattening. At higher H\,{\sc i} gas mass surface densities, the spectrum becomes more concave (lower value of $\beta$), as one would expect if ionisation losses play a role. The time-scale for ionisation losses is \citep{Murphy2009}:
\begin{equation}
    t_{\rm ion}=4.1\times 10^9 \left ( \frac{n}{\rm cm^{-3}}\right )^{-1} \left ( \frac{E}{\rm GeV}\right ) \left [ 3\ln \left ( \frac{E}{\rm GeV}\right ) + 42.5 \right ]^{-1}\,\rm yr. 
\end{equation}
Hence, in order to be an important process to flatten the radio continuum spectrum at low frequencies, this time-scale has to be shorter than the synchrotron loss time-scale of approximately 100\,Myr \citep{2023A&A...672A..21H}. For our cosmic-ray electrons with $E\approx 0.5\,\rm GeV$ at the lowest frequency, we thus expect $t_{\rm ion}\approx 100\,\rm Myr$ for neutral gas densities of $n\approx 0.5\,\rm cm^{3}$. This is the case for mass surface densities of $\Sigma_{\rm gas}=2.5$--$5\,\rm M_{\odot}\,pc^{-2}$, assuming a gas scale height of 100\,pc \citep{Cox2005,2015MNRAS.449.3879B}. This means the suggested correlation between non-thermal spectral curvature and H\,{\sc i} gas mass surface density can indeed be ascribed to cosmic-ray ionization losses that affect the low-energy cosmic-ray electrons the most. In the absence of a spectral model for ionisation losses, we leave more rigorous fitting of the spectra to future work. For now, we cannot rule out the effect free--free absorption in addition to ionisation losses. Hence, we conclude that the low-frequency flattening observed in M\,51 is probably caused by a combination of ionisation losses and free--free absorption. 

Ionisation of the molecular component of the interstellar medium should also cause CR energy loss. We repeated the same analysis shown in Fig. \ref{plot_HI_EM} to test the correlation between the spectral curvature and the total gas mass density. Contrary to our expectation, they were not found to be correlated ($\rho_{\rm s}=-0.07$). A possible reason is that the cosmic rays (at least in the GeV-range) do not enter molecular clouds, due to the strong and turbulent magnetic field \citep{2013A&A...552A..19T}.

\section{Summary and conclusions}
\label{section_discussion}

Synchrotron spectra at low radio frequencies are shaped by several processes such as cosmic-ray ionisation losses and free--free absorption that suppress the emission. The cosmic-ray energy loss from ionisation is nearly independent of energy (only logarithmic dependence) so it proportionally affects low-energy cosmic-ray electrons the most, resulting in spectral flattening. Free--free absorption is also strongly dependent on frequency with frequencies below 300\,MHz (depending on EM) being more affected. This affects our ability to interpret low-frequency continuum emission as an extinction-free star formation tracer. On the other hand, strong free--free absorption would allow us to measure the EM which itself is, derived from radio spectra, an extinction-free star-formation tracer. In order to study these effects we compiled data of the nearby grand-design spiral galaxy M\,51 at nine different frequency bands between 54 and 8350\,MHz of which we used eight for the analysis. We present here, for the first time, new observation with the GMRT at 240\,MHz. We calculated the contribution of the thermal radio continuum emission based on the H\,$\upalpha$ map corrected for extinction using 24\,$\upmu\rm m$ mid-infrared data. This contribution was subtracted from the radio maps so we studied only the non-thermal emission. We also used new integral field unit spectroscopy data from Metal-THINGS in order to measure extinction-corrected H\,$\upalpha$ intensities which we use as an independent estimate of EM.

First, we fit a polynomial function to the logarithmic intensities to determine the non-thermal radio spectral index and spectral curvature as shown in Fig.\,\ref{image_SI_SC}. The non-thermal radio spectral index at 1\,GHz is relatively flat in the spiral arms with $\alpha_0\approx -0.6$, in agreement with the injection spectral index. The spectral curvature is negative throughout the galaxy with more negative values in the spiral arms. This means the spectrum is concave where intensities both at low and high frequencies are suppressed. This is the expected behaviour for low-energy losses and absorption effects as well as strong synchrotron and inverse Compton radiation losses at higher frequencies resulting in spectral ageing. Next, we calculated the low- (54--144\,MHz) and high-frequency (1370--4850\,MHz) non-thermal two-point radio spectral indices separately as shown in Fig.\,\ref{image_SI}. 
The low-frequency radio spectral index is very flat and even positive in the spiral arms with $\alpha_{\rm low}$ between $-0.5$ and $0.2$, showing the presence of inverted radio continuum spectra. This clearly hints at either low frequency absorption or cosmic-ray ionisation losses. In contrast, the high-frequency radio spectra are fairly steep with values of $\alpha_{\rm high}$ between $-1.5$ and $-0.7$. The differences throughout the galaxy can be explained by spectral ageing. 

Next we analysed the spatially resolved spectral index in more detail, starting with the effect of free--free absorption. In Fig.\,\ref{plot_SEDs} we plotted the integrated spectra of the spiral arms, inter-arm and core regions. We find that the spectra can be well fitted with a free--free absorption model \citep{1997MNRAS.291..517W}. At 54\,MHz the deviation from a power-law spectrum is most apparent. 

Comparing the spiral arms with the inter-arm regions, we find that in the spiral arms the turnover occurs at higher frequencies, which implies stronger free--free absorption. This is confirmed by the values of the EM obtained by fitting the free--free absorption model. On the other hand, the difference in EM between spiral arm and inter-arm regions is not statistically significant, suggesting that free--free absorption might not be the only effect. 

We then fitted the free--free absorption model to our spatially resolved data. We obtain point-by-point maps of the EM and the radio spectral index $\eta$ (Fig.\,\ref{image_thermabs}). We compare the values of EM obtained from the fit to the values estimated from the extinction-corrected H\,$\upalpha$ measurements. We do not see a correlation between the EM's calculated from different tracers (Fig.\,\ref{plot_Ha_EM}). The lack of correlation, may be hence explained by the very different volume filling factors or it is possible that the free--free absorption is not the only process which causes the low frequency flattening. 
Therefore, we also investigated the possible influence of cosmic-ray ionisation losses on the radio continuum spectrum. To this end, we study the correlation between non-thermal spectral curvature and gas mass surface density. We find a weak correlation between spectral curvature and  H\,{\sc i} gas mass surface density (Fig.\,\ref{plot_HI_EM}). This leads us to believe that the main process responsible for the low-frequency flattening is more dominant in the regions with neutral H\,{\sc i} gas. In those regions the cosmic rays lose energy due to ionisation. Because ionisation losses are nearly frequency-independent, they are more apparent at low frequencies where we observe emission from low energy cosmic-ray electrons.  
We therefore conclude that the low frequency flattening in M\,51 is most likely caused by a combination of free--free absorption and ionisation losses.

This is the first time that we observe a low-frequency spectral flattening in an average spiral galaxy outside of the Milky Way. Previously, this effect has only been observed in starburst galaxies \citep{2013A&A...555A..23A, 2015MNRAS.449.3879B, 2016A&A...593A..86V, 2017A&A...608A..29A}. Now we were able to spatially locate the region of the galaxy where the spectra flattens at low frequencies and find that those areas generally have a higher density of H\,{\sc i} gas. Curiously, we do not find significant correlation between spectral curvature and total gas mass density, suggesting that the cosmic-ray ionisation losses are in the atomic gas phase but not in the molecular gas phase. Possibly because the cosmic rays do not enter molecular clouds due to the magnetic field \citep{2013A&A...552A..19T}. We also want to consider what our results mean for using sub-GHz radio continuum observations as a SFR tracer. Even using low frequencies of around 150 MHz (LOFAR HBA) as a SFR tracer requires an understanding of the losses we investigate here \citep[see e.g.][]{2013A&A...555A..23A, 2018A&A...619A..36C}. The forthcoming surveys with LOFAR at very low frequencies and high angular resolution will enable further studies in other nearby galaxies and allow us to investigate the influence of these effects in more detail.

\acknowledgements{
We thank the two anonymous referees for detailed comments which significantly improved the paper and especially made it more understandable for the reader.
LG and MB acknowledge funding by the Deutsche Forschungsgemeinschaft (DFG, German Research Foundation) under Germany's Excellence Strategy -- EXC 2121 ``Quantum Universe'' -- 390833306. BA, DJB and MS acknowledge funding from the German Science Foundation DFG, via the Collaborative Research Center SFB1491 ‘Cosmic Interacting Matters -- From Source to Signal’. FdG acknowledges support from the ERC Consolidator Grant ULU 101086378. LOFAR \citep{2013A&A...556A...2V} is the Low Frequency Array designed and constructed by ASTRON. It has observing, data processing, and data storage facilities in several countries, which are owned by various parties (each with their own funding sources), and that are collectively operated by the ILT foundation under a joint scientific policy. The ILT resources have benefited from the following recent major funding sources: CNRS-INSU, Observatoire de Paris and Université d'Orl\'eans, France; BMBF, MIWF-NRW, MPG, Germany; Science Foundation Ireland (SFI), Department of Business, Enterprise and Innovation (DBEI), Ireland; NWO, The Netherlands; The Science and Technology Facilities Council, UK; Ministry of Science and Higher Education, Poland; The Istituto Nazionale di Astrofisica (INAF), Italy. This research made use of the Dutch national e-infrastructure with support of the SURF Cooperative (e-infra 180169) and the LOFAR e-infra group. The J\"ulich LOFAR Long Term Archive and the German LOFAR network are both coordinated and operated by the J\"ulich Supercomputing Centre (JSC), and computing resources on the supercomputer JUWELS at JSC were provided by the Gauss Centre for Supercomputing e.V. (grant CHTB00) through the John von Neumann Institute for Computing (NIC).  This research made use of the University of Hertfordshire high-performance computing facility and the LOFAR-UK computing facility located at the University of Hertfordshire and supported by STFC [ST/P000096/1], and of the Italian LOFAR IT computing infrastructure supported and operated by INAF, and by the Physics Department of Turin university (under an agreement with Consorzio Interuniversitario per la Fisica Spaziale) at the C3S Supercomputing Centre, Italy. The research leading to these results has received funding from the European Research Council under the European Union's Seventh Framework Programme (FP/2007-2013) / ERC Advanced Grant RADIOLIFE-320745. The Westerbork Synthesis Radio Telescope is operated by ASTRON (Netherlands Foundation for Research in Astronomy) with support from the Netherlands Foundation for Scientific Research (NWO). This research made use of the Python Kapteyn Package \citep{KapteynPackage}.}

\bibliography{bibtex}{}
\bibliographystyle{aa}
\appendix

\onecolumn 
\section{Thermal subtraction}
\label{a:therm_subs}

\begin{figure}[!ht]
    \caption{Intermediate data products in the subtraction of thermal emission in M\,51. }
    \centering
    \begin{subfigure}[t]{0.48\textwidth}
        \centering
        \includegraphics[width=0.95\linewidth]{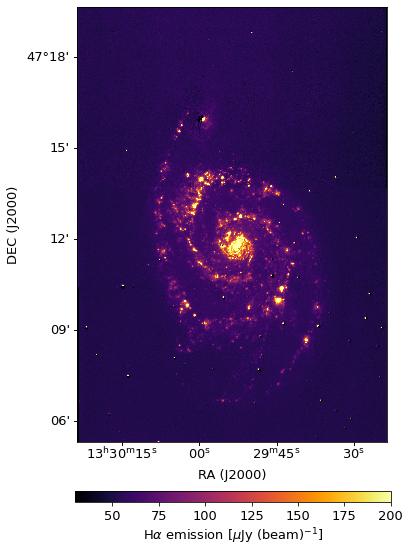}
        \label{fig:Ha_appendix}
        \caption{Initial continuum-subtracted H$\upalpha$ map obtained with the Kitt Peak National Observatory 2.1-m telescope using the narrow-band H$\upalpha$-filter KP1563 \citep{2003PASP..115..928K}. The angular resolution is $1\farcs 35 \times 1\farcs 35$ and the rms noise is $\approx$$50\,\upmu\rm Jy\, beam^{-1}$. The map was downloaded from the ancillary data at the SINGS webpage.\footnote{\url{http://irsa.ipac.caltech.edu/data/SPITZER/SINGS/}}}
    \end{subfigure}\hfill
    \begin{subfigure}[t]{0.48\textwidth}
        \centering
        \includegraphics[width=0.95\linewidth]{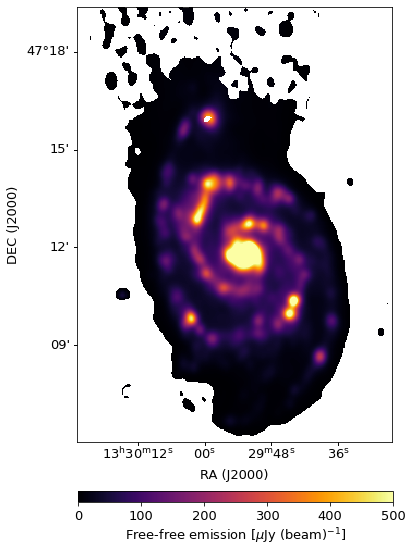}
        \label{fig:ff_appendix}
        \caption{Extinction corrected prediction of the free--free emission at 1370\,MHz in M51. This emission was scaled to the appropriate frequency with the spectral index index $\alpha=0.1$. The map shown here was subtracted from the radio maps to get the non-thermal emission. The beam size is $17 \farcs 5 \times 15\farcs 0$.}
    \end{subfigure}
    \label{fig:subtraction_appendix}
\end{figure}

\newpage
\section{Point spectra}
\label{a:point}

\begin{figure}[!h]
    \caption{Individual point spectra. }
    
    \begin{subfigure}[h]{0.48\textwidth}
        \centering
        \includegraphics[width=0.95\linewidth]{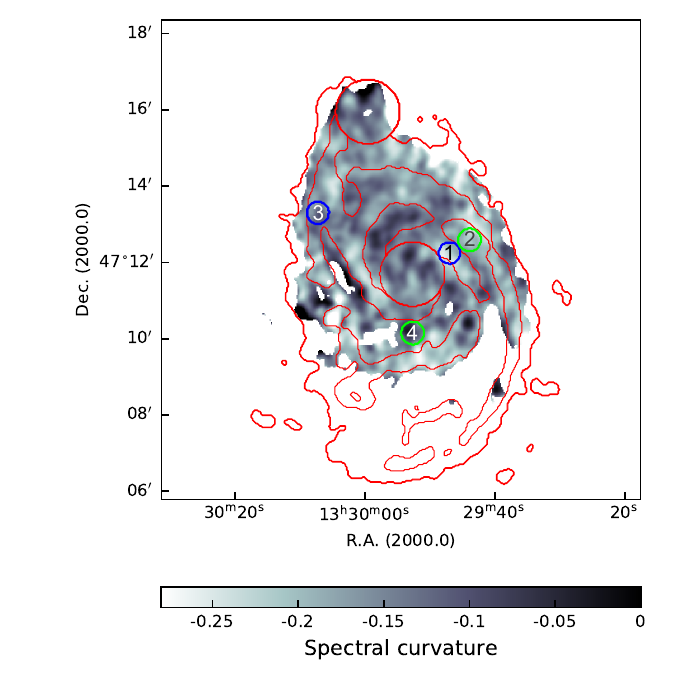}
        
        \caption{The location of individual spectra from spiral arms (interarm regions) marked by green (blue) circles on top of the spectral curvature map. The size of the circles corresponds to the beam size. Red contours at $8$ and $25\,\rm M_{\odot}\,pc^{-2}$ define the borders of the spiral arm and inter-arm regions, respectively (Sect. \ref{section_spiral_inter-arm}).  }
        \label{fig:locations_appendix}
    \end{subfigure}\hfill
    \begin{subfigure}[h]{0.48\textwidth}
        \centering
        \includegraphics[width=0.95\linewidth]{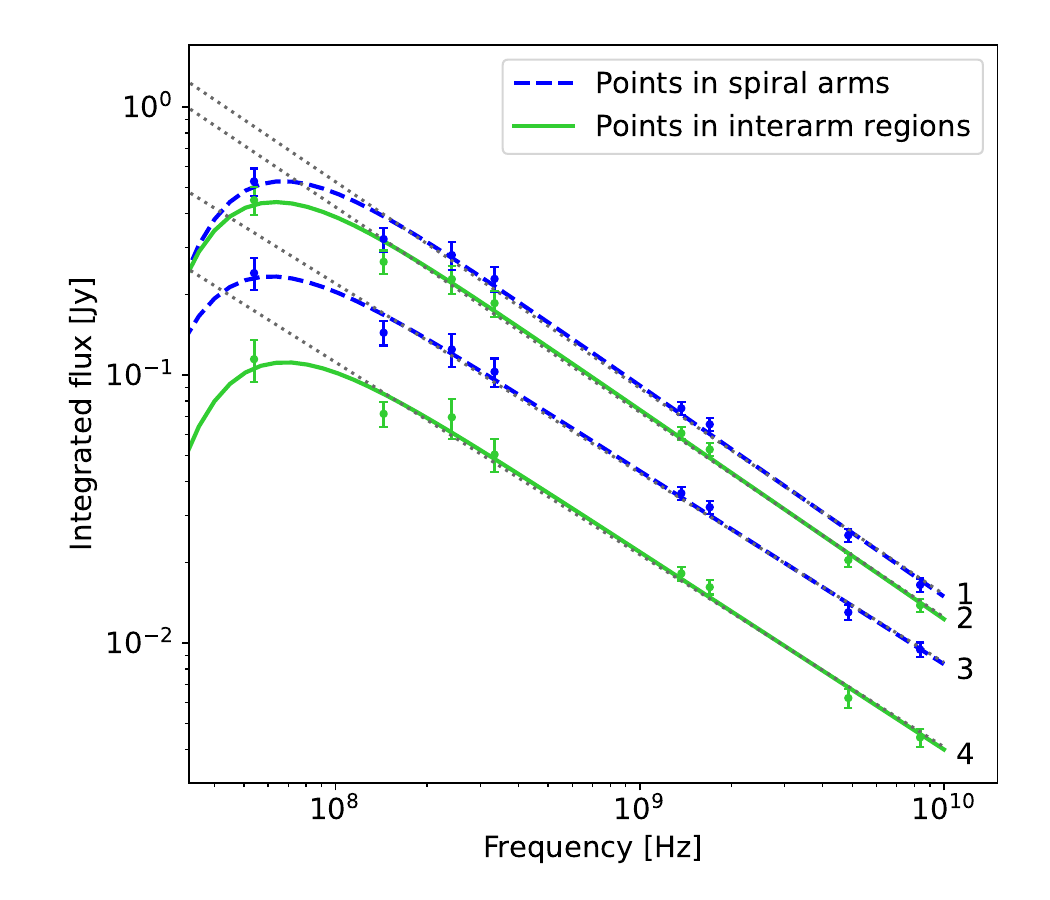}
        \label{plot_a_point}
        \caption{Non-thermal radio continuum spectra for representative beam-sized regions marked in Fig. \ref{fig:locations_appendix}. Each spectrum is labeled with its corresponding number. Dotted lines show the power-law fits for frequencies $\geq$144\,MHz. The individual spectra show the same trends as the integrated spectra in Fig. \ref{plot_SEDs}. }
    \end{subfigure}
    \label{fig:point_appendix}
\end{figure}

\end{document}